\documentclass[%
reprint,
superscriptaddress,
showpacs,
amsmath,amssymb,
aps,
pra,
Ifloatfix,
citeautoscript,
]{revtex4-2}
\usepackage{CJK}

\usepackage[english]{babel}
\usepackage{graphicx}
\usepackage{epstopdf}
\usepackage{mathrsfs}
\usepackage[capitalise]{cleveref}
\usepackage{mathtools}
\usepackage{subcaption}
\usepackage{float}
\usepackage{tensor}
\usepackage{siunitx}
\usepackage{dcolumn}
\usepackage{bm}
\usepackage[normalem]{ulem}
\usepackage{color}
\usepackage[inline]{enumitem}

\providecommand{\diff}[1]{\ensuremath{\mathrm{d}{#1}}}
\providecommand{\rbracket}[1]{\ensuremath{\left( #1 \right)}}

\providecommand{\sbracket}[1]{\ensuremath{\left[ #1 \right]}}

\providecommand{\cbracket}[1]{\ensuremath{\left\{ #1 \right\}}}

\usepackage{tensor}
\usepackage{braket}

\begin{document}

\begin{CJK*}{UTF8}{}
				
\title{Quantum optical levitation of a mirror}
\CJKfamily{bsmi}
\author{C. T. Marco Ho \CJKkern(何宗泰)}
\affiliation{Centre for Quantum Computation and Communication Technology, School of Mathematics and Physics, University of Queensland, Brisbane, Queensland 4072, Australia}

\author{Ryan J. Marshman}%
\email{r.marshman@uq.edu.au}
\affiliation{Centre for Quantum Computation and Communication Technology, School of Mathematics and Physics, University of Queensland, Brisbane, Queensland 4072, Australia}

\author{Robert B. Mann}
\email{rbmann@uwaterloo.ca}
\affiliation{Centre for Quantum Computation and Communication Technology, School of Mathematics and Physics, University of Queensland, Brisbane, Queensland 4072, Australia}
\affiliation{Department of Physics and Astronomy, University of Waterloo, Waterloo, Ontario N2L 3G1, Canada}
\affiliation{Perimeter Institute, 31 Caroline Street North, Waterloo, Ontario N2L 2Y5, Canada}
		
\author{Timothy C. Ralph}%
\email{ralph@physics.uq.edu.au}
\affiliation{Centre for Quantum Computation and Communication Technology, School of Mathematics and Physics, University of Queensland, Brisbane, Queensland 4072, Australia}
		
		\date{\today}
		
		\begin{abstract}
		While the levitating mirror has seen renewed interest lately, relatively little is known about its quantum behaviour. In this paper we present a quantum theory of a one dimensional levitating mirror. The mirror forms a part of a Fabry-P\'{e}rot cavity where the circulating intracavity field supports the mirror through radiation pressure alone. We find a blue and red detuned steady-state of which only the blue detuned solution with damping on the mirror and cavity is stable. We find strong entanglement (15-20 ebits) between the mirror output and cavity output and squeezing in the mirror output.
		\end{abstract}
		\maketitle
\end{CJK*}

	
	\section{Introduction \label{sec:introduction}}
	Levitation by light is an accessible stage where we can see the push and pull between gravity and quantum physics. Optomechanical systems such as pendula, in which gravity provides part of the restoring force, have proven to be extremely versatile, with applications ranging from the generation of squeezed light \cite{marino_classical_2010,safavi-naeini_squeezed_2013}, to laser cooling of harmonic oscillators \cite{groblacher_demonstration_2009}, to precision metrology \cite{tsang_quantum_2013,carmon_temporal_2005}, with mirror masses from the nanoscale \cite{sun_femtogram_2012} to the kiloscale \cite{corbitt_measurement_2006,corbitt_review:_2004}. Such macroscopic optomechanical systems have opened up the possibility of testing quantum-gravity interaction models \cite{grosardt_optomechanical_2016,gan_optomechanical_2016}. 
	
	The ultimate such system, often invoked as a gedanken experiment, is when a mirror is solely suspended by radiation pressure. Such levitating systems have been proposed as a way to reduce noise and decoherence from unwanted coupling \cite{arvanitaki_detecting_2013,libbrecht_toward_2004}. As a low-dissipative \emph{and} macroscopic optomechanical system, the  levitating mirror has been of renewed interest lately, with a tripod \cite{guccione_scattering-free_2013} and a double Fabry-P\'{e}rot cavity \cite{michimura_optical_2017} both recently proposed as possible systems. Mirrors and nano-particles supported by optical tweezers (with gravity neglected) have been analysed quantum mechanically \cite{SIN10,ISA10,CHA10} However, there has not yet been a full quantum optomechanical analysis of a levitating mirror for which gravity and radiation pressure are the only restoring forces. 
	
	In this paper, we derive the quantum theory for a one dimensional levitating mirror and examine its stability and dynamics in an experimentally relevant regime. We find significant damping of the mirror is required to achieve stability and strong entanglement is generated between the mirror and the driving field. We also consider the effective temperature due to both the driven mirror dynamics and absorption effects, and show that in practice this will make any entanglement difficult to observe in this configuration.

	\section{System Hamiltonian \label{sec:hamiltonian}}
\begin{figure}[htb]
		\centering
		\includegraphics*[width=0.7\columnwidth]{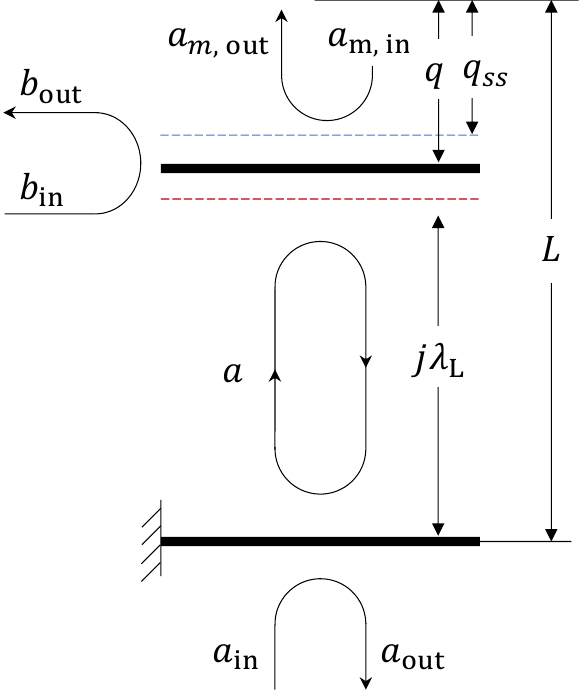}
		\caption{A visual representation of the system. The bottom mirror is fixed and forms a cavity with the free floating mirror above. The red and blue dashed lines indicate red ($\Delta>0$) and blue ($\Delta<0$) detuning respectively. The red detuned case is unstable. Here $L$, $q$ and $q_{ss}$ are the cavity rest length, current mirror position and steady state position respectively while $a$, $a_{in}$, $a_{out}$, $a_{m,in}$, $a_{m,out}$, $b_{in}$ and $b_{out}$ represent the operators for the cavity photons, lower mirror input-output photons, upper mirror input-output photons and upper mirror phonon input and output respectively.}
		\label{fig:bluedetuned}
	\end{figure}

	Let us consider a one dimensional Hamiltonian for a Fabry-P\'{e}rot cavity where the lower mirror is stationary and the upper mirror (henceforth, referred to as the mirror) is free to move along the cavity axis. We will couple a laser into the cavity, which will support the mirror by radiation pressure alone. $\hat{q}$ and $\hat{p}$ are the position and momentum operators of the mirror, measured from a resting length $L$ (see \cref{fig:bluedetuned}) and $\widehat{\Omega}_c (\hat{q})=\frac{j\pi c}{L-\hat{q}}$ is the position dependent resonant frequency of the cavity, with $j$ labeling the $j$th mode of the cavity to which the laser is coupled. $\hat{a}$ and $\hat{a}^\dagger$ are the annihilation and creation operators of the intracavity mode with the commutation relation $\sbracket{\hat{a},\hat{a}^\dagger}=1$ and $\sbracket{\hat{q},\hat{p}} = i\hbar$. In the  rotating frame of the laser frequency $\Omega_L$, the Hamiltonian is
	\begin{equation}
	H= \frac{\hat{p}^2}{2m}-mg\hat{q}+\hbar\rbracket{\widehat{\Omega}_c (\hat{q})-\Omega_L} \hat{a}^\dagger \hat{a}.
	\end{equation}
	We use input-output theory to probe the behaviour of the system.
	To model the coupling of the laser into the cavity, absorption of the cavity field by the mirror and the interaction of the system with the environment, we will assume that the cavity and the mirror are individually coupled to Markovian baths \cite{bowen_quantum_2015}. The mirror has almost perfect reflectivity while the lower stationary mirror is the input-output port to which the laser is coupled. This leads to the operator equations of motion \cite{bowen_quantum_2015,gardiner_input_1985} for the cavity and the mirror
	\begin{align}
	\dot{\hat{a}} &= -\sbracket{\frac{\kappa}{2} + i \rbracket{\widehat{\Omega}_c (\hat{q})) -\Omega_L}}\hat{a} + \sqrt{\kappa} \hat{a}_\text{in} + \sqrt{\eta}\hat{a}_{m, in}\label{eq:adot}\\
	\dot{\hat{q}} &= \frac{\hat{p}}{m} 
	\\
	\dot{\hat{p}} & =mg - \frac{\hbar \widehat{\Omega}^2_c (\hat{q})}{j \pi c} \hat{a}^\dagger \hat{a}- \frac{\Gamma}{2} \hat{p} + \sqrt{\Gamma} \hat{p}_\text{in},  
	\end{align}
	where $\hat{p}_\text{in}$ is the input momentum due to the mirror bath and $\Gamma$ is the damping due to the mirror's bath. $\hat{a}_\text{in}$ is the input due to the cavity's bath and the $\kappa$ is the damping rate due to the cavity's bath. Similarly $\hat{a}_\text{m, in}$ is the input due to the mirror absorption mechanism and the $\eta$ is the damping rate due to the mirror's bath which will take as zero until we consider thermalisation in Section \ref{sec:temperature}. Hence, we assume initially that the cavity damping $\kappa$ is due solely to the reflectivity of the fixed, lower mirror while the mirror damping $\Gamma$ is ultimately due to radiative damping.
	
	We will assume that these operators $\mathcal{\widehat{O}}$ can be written as $\mathcal{\widehat{O}}= \mathcal{O}_\text{SS} + \delta\mathcal{\widehat{O}}$ composed of a steady-state solution $\Braket{\mathcal{\widehat{O}}}= \mathcal{O}_\text{SS}$ and a fluctuating component, $\delta \mathcal{\widehat{O}}$, that is small, $ \Braket{\delta \mathcal{\widehat{O}} \delta \mathcal{\widehat{O}}} \ll \mathcal{\widehat{O}}_\text{SS}^2$, and has zero expectation, $\Braket{\delta \mathcal{\widehat{O}}}=0$. To derive semi-classical steady-state equations, we take the expectation value of the equations and set $\Braket{\dot{\hat{a}}} = \Braket{\dot{\hat{q}}} = \Braket{\dot{\hat{p}}}=0$. We will also require that $q_\text{in,SS} =p_\text{in,SS}=0$ as there is no coherent force on the mirror. In contrast $\hat{a}_\text{in,SS}$ is non-zero as the bath input is a coherent laser beam that provides the power to levitate the mirror. Defining the steady-state cavity number \footnote{Due to the requirement that the linearised part is small, $\Braket{\hat{a}^\dagger \hat{a}} = \Braket{\hat{a}^\dagger}\Braket{\hat{a}} + \Braket{\delta \hat{a}^\dagger \delta \hat{a}} \approx \Braket{\hat{a}^\dagger}\Braket{\hat{a}}$. } $N_c \equiv \Braket{\hat{a}^\dagger \hat{a}} \approx \Braket{\hat{a}^\dagger}\Braket{\hat{a}}$ and the input photon rate $N_\text{in} = \Braket{\hat{a}^\dagger_\text{in}\hat{a}_\text{in}}\approx \Braket{\hat{a}^\dagger_\text{in}}\Braket{\hat{a}_\text{in}}$, we find that there are two steady-state solutions, $q_1, N_{c,1}$ and $q_2, N_{c,2}$.
	
	We define  $\widetilde{N}_\text{in} \equiv N_\text{in}/mg$ such that the steady-state $q_i$ and $\widetilde{N}_{c,i}$ can be written as $q_i = q_i(\widetilde{N}_\text{in},\Omega_L, j, L, \kappa)$ and $\widetilde{N}_{c,i} = \widetilde{N}_{c,i} (\widetilde{N}_\text{in},\Omega_L, j, L, \kappa)$. 
	
	\section{Conditions of parameters and detuning \label{sec:detuning}}
	As the laser enters the cavity it should couple with the mode $j$ with the closest frequency or smallest `detuning' $\Delta \equiv\frac{j\pi c}{L-q} -\Omega_L$. We will thus approximate $j =\text{Round}\rbracket{\frac{L\Omega_L}{\pi c}}$. In particular, to ensure that we are really only addressing one mode, we require that $\Delta \ll \frac{\pi c}{L-q}$, which means that the detuning is small with respect to the frequency spacing $\frac{\pi c}{L-q}$ between the modes in the cavity. Furthermore, the damping rate $\kappa$ for a good cavity must be much smaller than the frequencies in play such as $\Omega_L$ and $\Omega_c$; therefore, we also require that $\frac{\kappa}{\Omega_L}\ll 1$.

	In optomechanical systems the detuning can either `cool' or `heat' the oscillator. Downward phonon (mirror excitons) number transitions are enhanced when you have red detuning $(\Delta >0)$ while for blue detuning $(\Delta <0)$ upward transitions are enhanced \cite{bowen_quantum_2015}. The cooling from red detuning is a problem for our floating mirror, as it can continue to lose energy by falling lower. This suggests that only blue detuned solutions with sufficient damping on the mirror can be stable. Given the importance of the detuning, we need to determine the detuning of the two steady-state solutions. To do this, we need to have reasonable estimates of the steady state parameters.  Let us consider the simplest case of a black object that is levitated by a laser that must be supported by laser power $P = \hbar N_\text{in} \Omega_L = mgc$,
where $N_\text{in}$ has dimensions of photons per second. This gives us a rough idea of how large $N_\text{in}$ should be and suggests defining  a dimensionless power $\widetilde{P} = \frac{P}{mgc} = \frac{\hbar \widetilde{N}_\text{in} \Omega_L}{c}$. With this definition we can show that the parameter dependence of the detuning is $\Delta = \Delta ( \widetilde{P}, \Omega_L,L,\kappa)$. If we set $P\approx mgc$ and assume the conditions specified above we find that $\Delta_1<0$ and $\Delta_2 >0$, that is, the first solution is blue detuned while the second is red detuned \footnote{See appendix for more details.}. Later we will vary $P$, but the conclusions about the detuning of steady-state solutions will still hold.

	In our system, the mirror can be said to be floating on a `bed' of photons that act like a spring. The mirror is not naturally a harmonic oscillator; only the perturbations of the mirror around the steady state act like a harmonic oscillator. Recalling our previous definition for the linearisation $\mathcal{\widehat{O}} = \Braket{\mathcal{\widehat{O}}} + \delta \mathcal{\widehat{O}}$, for small variations of $\hat{p}$, $\hat{q}$, $\hat{a}$ and $\hat{a}^\dagger$, we define
	\begin{align}
	\delta \hat{q} &= \sqrt{\frac{\hbar}{2m\Omega_M}} \rbracket{\delta \hat{b} + \delta \hat{b}^\dagger},\label{eq:position perturbation}\\
	\delta \hat{p} & = i \sqrt{\frac{\hbar m \Omega_M}{2}} \rbracket{\delta \hat{b}^\dagger - \delta \hat{b}}\label{eq:momentum perturbation},
	\end{align}
	where the mirror frequency $ \Omega_M^2  = \frac{2\hbar \Omega_c^3 N_c}{m(j\pi c)^2}$ and the coupling strength $g_C= \frac{ \Omega_c^2}{j\pi c}\sqrt{\frac{\hbar}{2m\Omega_M}} \alpha$. Assuming that $\alpha$ is real (which is an arbitrary choice of phase reference), our linearised Hamiltonian with the rotating wave approximation is
	\begin{equation}
	\delta H = \hbar \Delta \delta \hat{a}^\dagger \delta \hat{a}+ \hbar \Omega_M \delta \hat{b}^\dagger \delta \hat{b} +  \hbar g_C \rbracket{ \delta \hat{a} +  \delta \hat{a}^\dagger} \rbracket{\delta \hat{b}+\delta \hat{b}^\dagger}. \label{eq:mirrorlinearisedhamiltonian}
	\end{equation}
	The perturbations of the mirror ($\delta \hat{b}$) act like a harmonic mechanical oscillator with a frequency $\Omega_M$ that couples to the perturbations of the intra-cavity field ($\delta \hat{a}$) with coupling strength $g_C$. 
	\Cref{eq:mirrorlinearisedhamiltonian} has the exact same form as a standard linearised optomechanical Hamiltonian.  However, in contrast to standard optomechanical systems where the Hamiltonian parameters are independent and freely variable,  the frequency for the mirror oscillator $\Omega_M$ and the coupling $g_C$ are dependent on the steady state solutions.
	
	Dropping the $\delta$s for clarity of notation, the linearised equations of motion are
	\begin{equation}
	\frac{\mathrm{d}}{\diff{t}}\begin{pmatrix}
	\hat{b}\\
	\hat{b}^\dagger\\
	\hat{a}\\
	\hat{a}^\dagger
	\end{pmatrix} = A \begin{pmatrix}
	\hat{b}\\
	\hat{b}^\dagger\\
	\hat{a}\\
	\hat{a}^\dagger
	\end{pmatrix}+\begin{pmatrix} 
	\sqrt{\Gamma} \hat{b}_\text{in}\\
	\sqrt{\Gamma} \hat{b}_\text{in}^\dagger\\
	\sqrt{\kappa} \hat{a}_\text{in}\\
	\sqrt{\kappa} \hat{a}_\text{in}^\dagger
	\end{pmatrix},\label{eq:timedomain}
	\end{equation}
	where
	\begin{equation}
	A = \begin{pmatrix}
	-\frac{\Gamma}{2} -i \Omega_M & 0& -i g_C& -ig_C\\
	0&  -\frac{\Gamma}{2} + i \Omega_M & i g_C & i g_C\\
	-i g_C & -i g_C & - \frac{\kappa}{2} -i \Delta& 0 \\
	i g_C & i g_C & 0 &  - \frac{\kappa}{2} +i \Delta 
	\end{pmatrix}.\label{eq:A}
	\end{equation}
	In contrast to the steady state solutions, quantities in the linearised fluctuation theory have slightly different parameter dependence. The detuning is $\Delta = \Delta ( \widetilde{P}, \Omega_L,L,\kappa)$, but the mechanical frequency and coupling have additional dependence on $g$: $\Omega_M = \Omega_M ( \widetilde{P}, \Omega_L,L,\kappa,g)$ and $g_C = g_C( \widetilde{P}, \Omega_L,L,\kappa,g)$. While we do not indicate dependence on $c$ and $\hbar$, those being constants of nature, we do indicate a dependence on $g$ as this might be a parameter that could be changed by accelerating the system or locating it at different heights. 
	
	\emph{Parameters}---As we will be considering a table top experiment, for this paper we will set $L=\SI{5}{\centi\meter}, g=\SI{9.81}{\metre\second^{-2}}, c = \SI{3e8}{\metre\second^{-1}},\lambda_L= \SI{1050}{\nano\metre}$. The simplest parameter for the experimentalist to change is the laser power, which is encapsulated by changes in $\widetilde{P}$. This has the added advantage of also visualising changes in the mass, as the steady state solutions are a function of $\widetilde {P}$ and not mass or laser power alone. 
	
	\section{Stability \label{sec:stability}}
\begin{figure}
		\includegraphics[width=0.8\columnwidth]{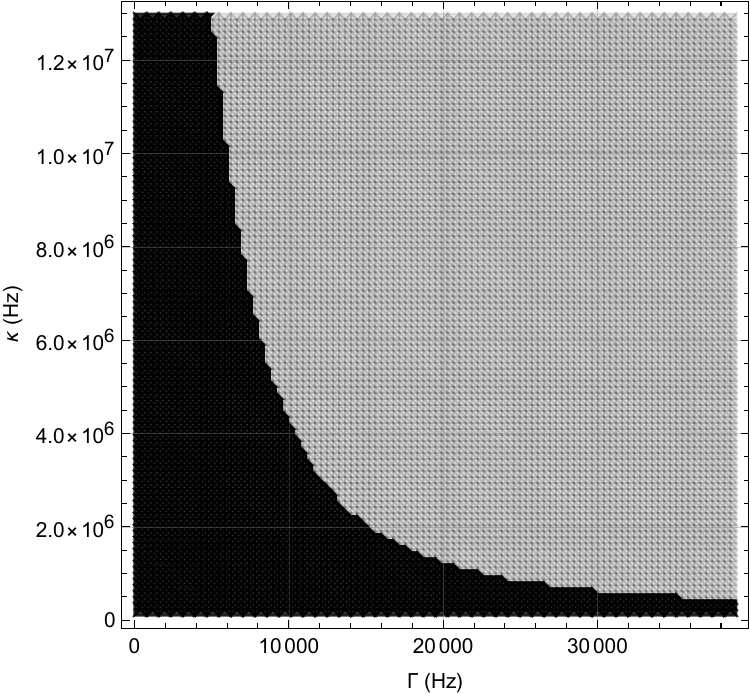}\
		\caption{Stability analysis with varying $\kappa$ and $\Gamma$. Light grey indicates stable regions while dark grey indicates unstable regions. Parameters are: $\widetilde {P} = 0.0017$, $L=\SI{5}{\centi\meter}$ and $\lambda_L= \SI{1050}{\nano\metre}$.}
		\label{stability}
	\end{figure}

	Being blue detuned does not ensure that the linearised equations are stable oscillations around the steady state. Therefore, we impose the condition that the real parts of the eigenvalues of $A$ are less than or equal to zero. This prevents the linearised solutions from exponential growth in time. With this criteria, we find, in agreement with our previous discussion, that the red detuned solution is always unstable while for sufficient damping in $\kappa$ and $\Gamma$ we can find a stable blue detuned solution (see \cref{stability}). We also note that we must have damping in both $\kappa$ and $\Gamma$ for a stable solution to exist. While we can control $\kappa$ by adjusting the reflectivity, we do not have a similarly straightforward method for tuning the mirror radiative damping. In principle radiative damping could be controlled by the physical properties of the mirror or by introducing a second cavity above the floating mirror to enhance radiation at particular frequencies. We will discuss this later in the paper.  To ensure stability, unless otherwise stated, we now only consider the blue detuned solution and we will set $\kappa = \SI{1.35e7}{\radian\per\second}$ and $\Gamma=\SI{1e4}{\radian\per\second}$.
	
	Interestingly for this set of parameters, the mechanical frequency $\Omega_M\sim O\left(10\right)$ is significantly less than all other fluctuation dynamics ($g_c,\left|\Delta\right|\gtrsim 10^{6}$) although, as we show later, the driven mechanical frequency is much higher as seen by the resonant frequency at which entanglement and squeezing occurs.
	
	\section{Frequency space solution to linearised equations of motion \label{sec:frequency space}}
	We want to find a solution to \cref{eq:timedomain} in the frequency domain. Note that we are using Fourier transforms of the conjugate, $\hat{a}^\dagger(\omega)$ instead of conjugates of the Fourier transform, $\hat{a}(\omega)^\dagger$. The two definitions are related by $\hat{a}^\dagger(\omega) = a(-\omega)^\dagger$. With these definitions the solution to the equations of motion in frequency space are $\rbracket{\hat{b},\hat{b}^\dagger,\hat{a},\hat{a}^\dagger} = T \rbracket{\sqrt{\Gamma} \hat{b}_\text{in},\sqrt{\Gamma} \hat{b}_\text{in}^\dagger,\sqrt{\kappa} \hat{a}_\text{in},\sqrt{\kappa} \hat{a}_\text{in}^\dagger}$ where $T= \rbracket{-i \omega I -A}^{-1}$.
	
	The input-output relations \cite{bowen_quantum_2015,gardiner_input_1985} are $\hat{a}_\text{out}(\omega) = \hat{a}_\text{in}(\omega)-\sqrt{\kappa}  \hat{a}(\omega)$ and $	\hat{b}_\text{out}(\omega) = \hat{b}_\text{in}(\omega)-\sqrt{\Gamma}  \hat{b}(\omega)$.	These relate input and system operators to measurable output operators. From these we define the position and momentum frequency quadratures to be $ \widehat{Q}_{\hat{a}_\text{out}} (\omega) = \frac{1}{\sqrt{2}}\rbracket{\hat{a}_\text{out} (\omega)+ \hat{a}_\text{out}^\dagger (\omega)}$ and $ \widehat{P}_{\hat{a}_\text{out}} (\omega) = \frac{-i}{\sqrt{2}}\rbracket{\hat{a}_\text{out} (\omega)- \hat{a}_\text{out}^\dagger (\omega)}$ respectively. We also define the quadratures in a similar way for $\hat{b}_\text{out}$. 
	
	\emph{Covariance matrix}---It is known that Hamiltonians that are bilinear in creation and annihilation operators preserve and create Gaussian states \cite{schumaker_quantum_1986,adesso_continuous_2014}.  Gaussian states are quantum states that are described by Gaussian Wigner functions which are fully characterised by their covariance matrices. As our linearised Hamiltonian is bilinear in operators, we will characterise our system using covariance matrices. Covariance matrices are usually defined from the quadrature operators. However, the frequency quadratures that we have used so far are only Hermitian when there is zero detuning. In the presence of detuning they are not directly measurable as they are not Hermitian. If we use homodyne detection, what we measure is the time domain quadratures $\widehat{Q}_{i_\text{out}}(t)$ and $\widehat{P}_{i_\text{out}}(t)$ where $i, j,\ldots \in \cbracket{a,b}$ indicate the mode. For this section we will suppress the \emph{out} subscript for legibility. To get the proper frequency quadratures, we mix-down the time domain quadratures with a cosine \cite{ralph_single-photon_2008} to get the symmetric, Hermitian quadrature $ \widehat{Q}^{C}_i(\omega) = \sqrt{\frac{2}{\pi}} \int \diff{t} \frac{\cos\rbracket{\omega t}}{\sqrt{2}} \widehat{Q}_i(t)=\frac{1}{\sqrt{2}}\rbracket{ \widehat{Q}_i(\omega) + \widehat{Q}_i(-\omega)}$ and the accompanying sine mix-down gives the antisymmetric, Hermitian quadrature, $\widehat{Q}^{S}_i(\omega) =\frac{-i}{\sqrt{2}} \rbracket{\widehat{Q}_i(\omega) - \widehat{Q}_i(-\omega)}$, with similar expressions for the momentum quadratures.
	To simplify notation, we define the vector of quadratures $\widehat{R}_i(\omega) =\left(\widehat{Q}^C_b,\widehat{P}^C_b,\widehat{Q}^S_b,\widehat{P}^S_b,\widehat{Q}^C_a,\widehat{P}^C_a,\widehat{Q}^S_a,\widehat{P}^S_a\right)$
	from which we define the real and symmetric covariance matrix
	\begin{equation}
	\sigma_{ij}(\omega) = \frac{1}{2}\Braket{\widehat{R}_i\widehat{R}_j + \widehat{R}_j\widehat{R}_i} - \Braket{\widehat{R}_i}\Braket{\widehat{R}_j}.
	\end{equation}
	Because the matrix is symmetric, the general form of the covariance matrix is given by,
	\begin{equation}
	\sigma = \begin{pmatrix}
	\sigma_b&\sigma_\text{upper}\\
	\sigma_\text{upper}^T & \sigma_a
	\end{pmatrix},
	\end{equation}
	where $\sigma_b$ and $\sigma_a$ are symmetric submatrices of the mirror and cavity respectively; their cross-correlations are given by $\sigma_\text{upper}$, see the appendix for more information. For a given $\omega$ and steady state parameters, the submatrices $\sigma_a$ and $\sigma_b	$ have off-diagonal terms. These off-diagonal terms indicate coupling between the symmetric and antisymmetric (cosine and sine) side-bands. While it is difficult to derive a closed-form expression, it can be shown that the two matrices can be independently diagonalised leading to two independent linear combinations of the symmetric and antisymmetric side-bands. 
	\begin{figure}[h]
	\centering
	\includegraphics[width=0.9\columnwidth]{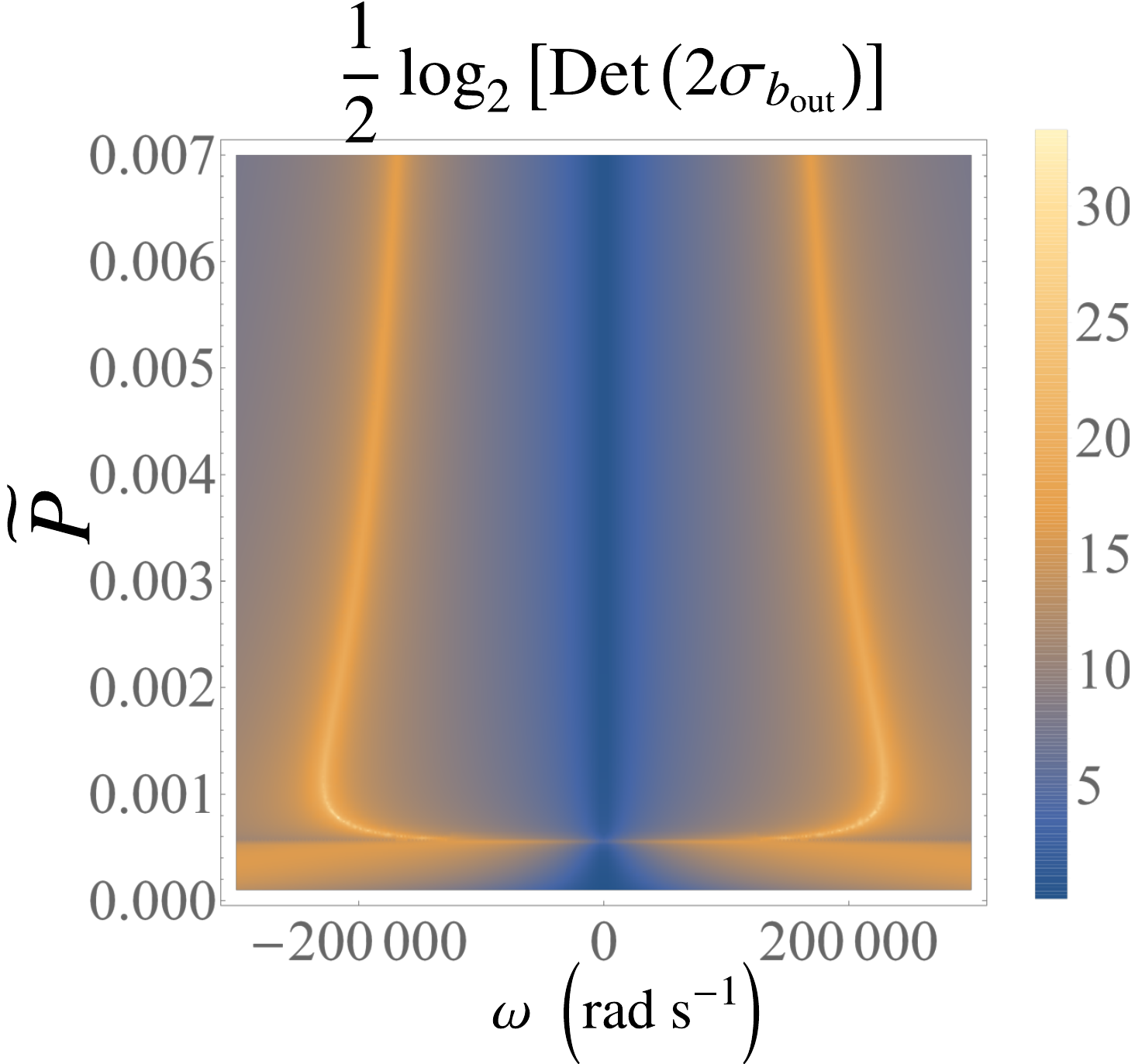}
	\caption{A plot of the entropy of entanglement between the output of the mirror and cavity at different side-band frequencies $\omega$ with respect to the laser frequency, as a function of dimensionless laser power, $\widetilde{P}$. Note that below $\widetilde{P}\approx0.00056$ the fluctuation analysis suggests the mirror is unstable. Parameters are: $L=\SI{5}{\centi\meter}$, $\lambda_L= \SI{1050}{\nano\metre}$, $\kappa = \SI{1.35e7}{\radian\per\second}$ and $\Gamma=\SI{1e4}{\radian\per\second}$.}
	\label{fig:entanglement}
	\end{figure}

	\begin{figure*}
		\centering
		\begin{subfigure}[l]{0.8\columnwidth}
			\includegraphics[width=\columnwidth]{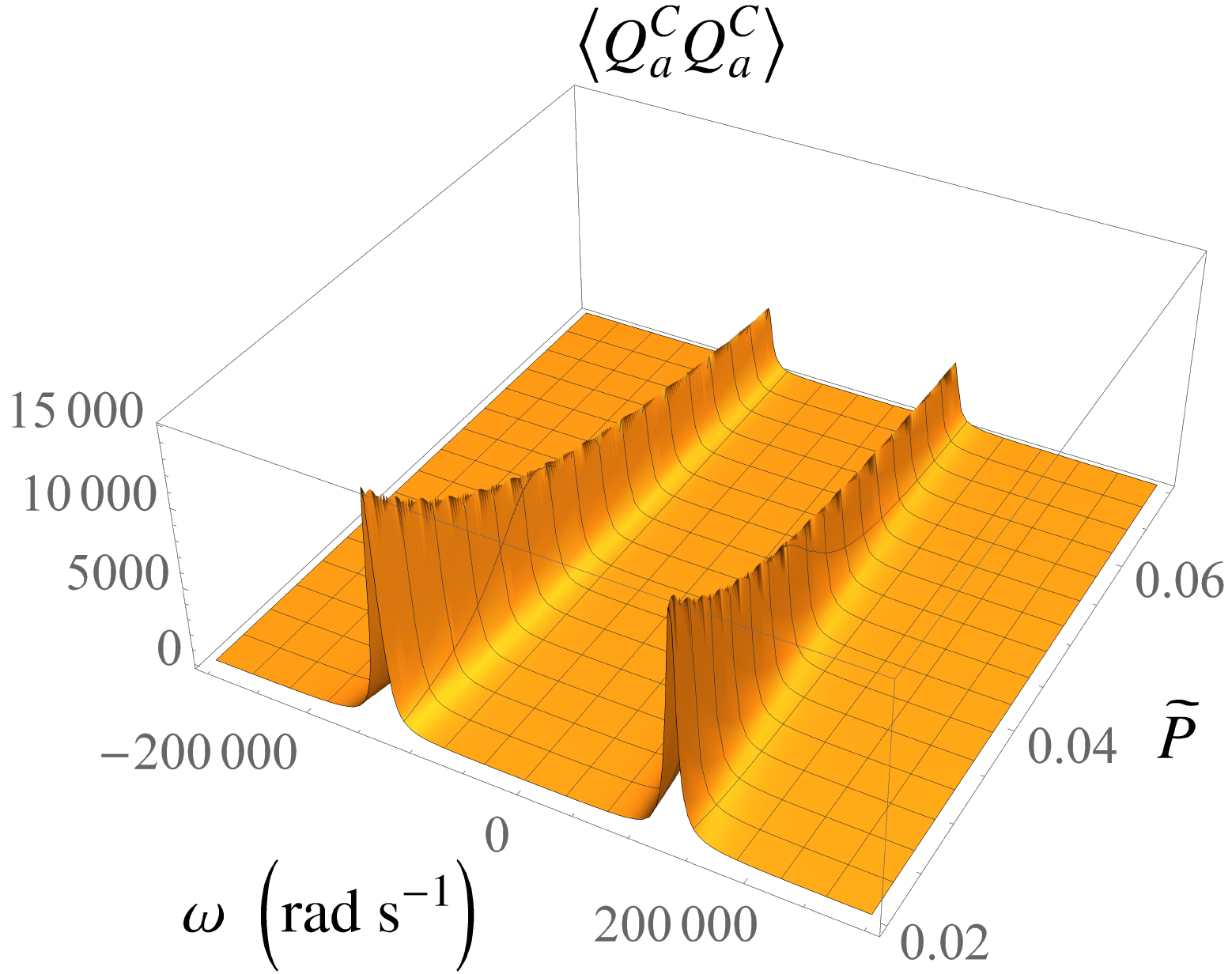}
			\caption{3D plot of the variance of $\widehat{Q}^{C/S}_a$. }
		\end{subfigure}
		\begin{subfigure}[l]{0.8\columnwidth}
			\includegraphics[width=\columnwidth]{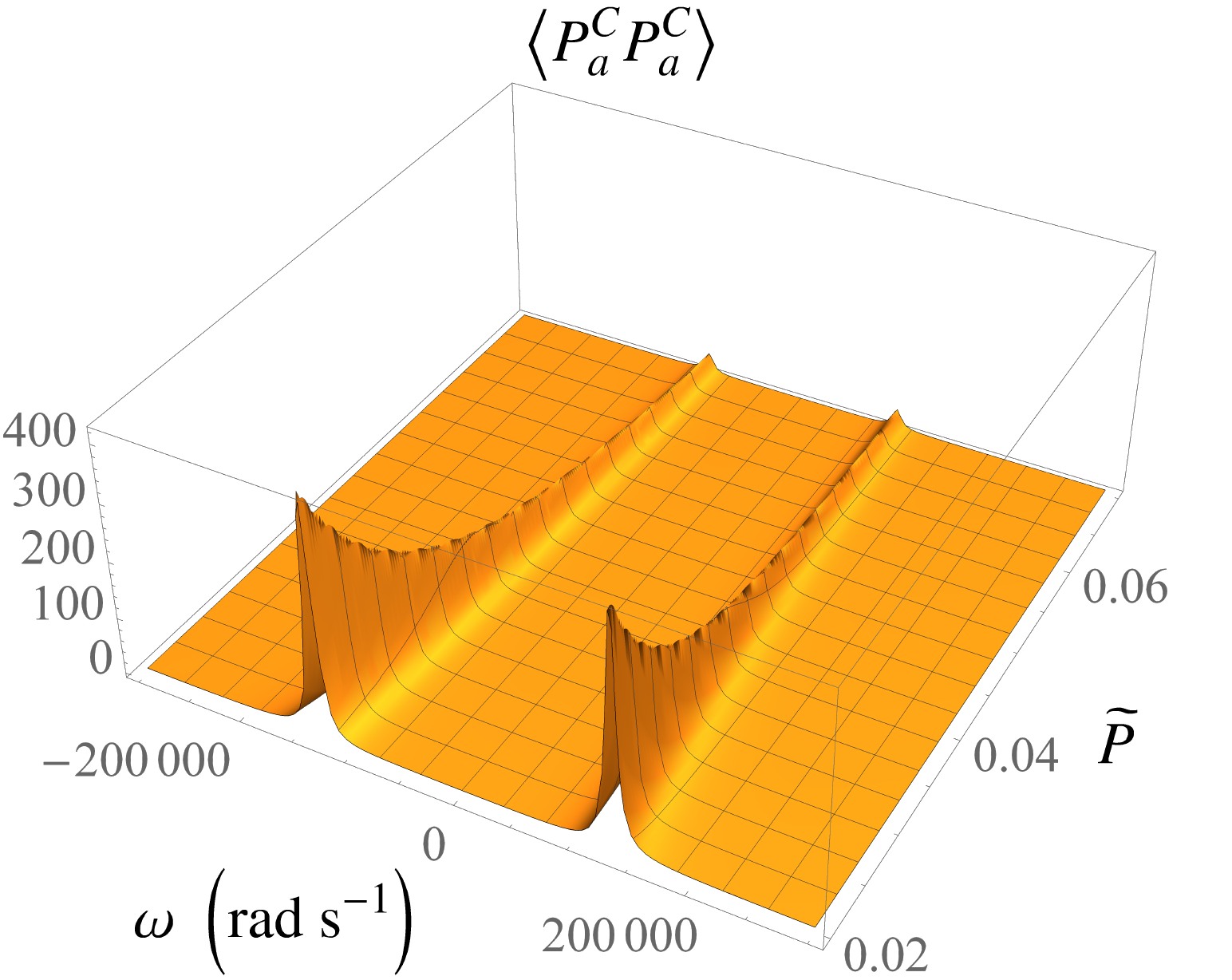}
			\caption{3D plot of the variance of $\widehat{P}^{C/S}_a$.}
		\end{subfigure}
		\par\medskip
		\begin{subfigure}[l]{0.8\columnwidth}
			\includegraphics[width=\columnwidth]{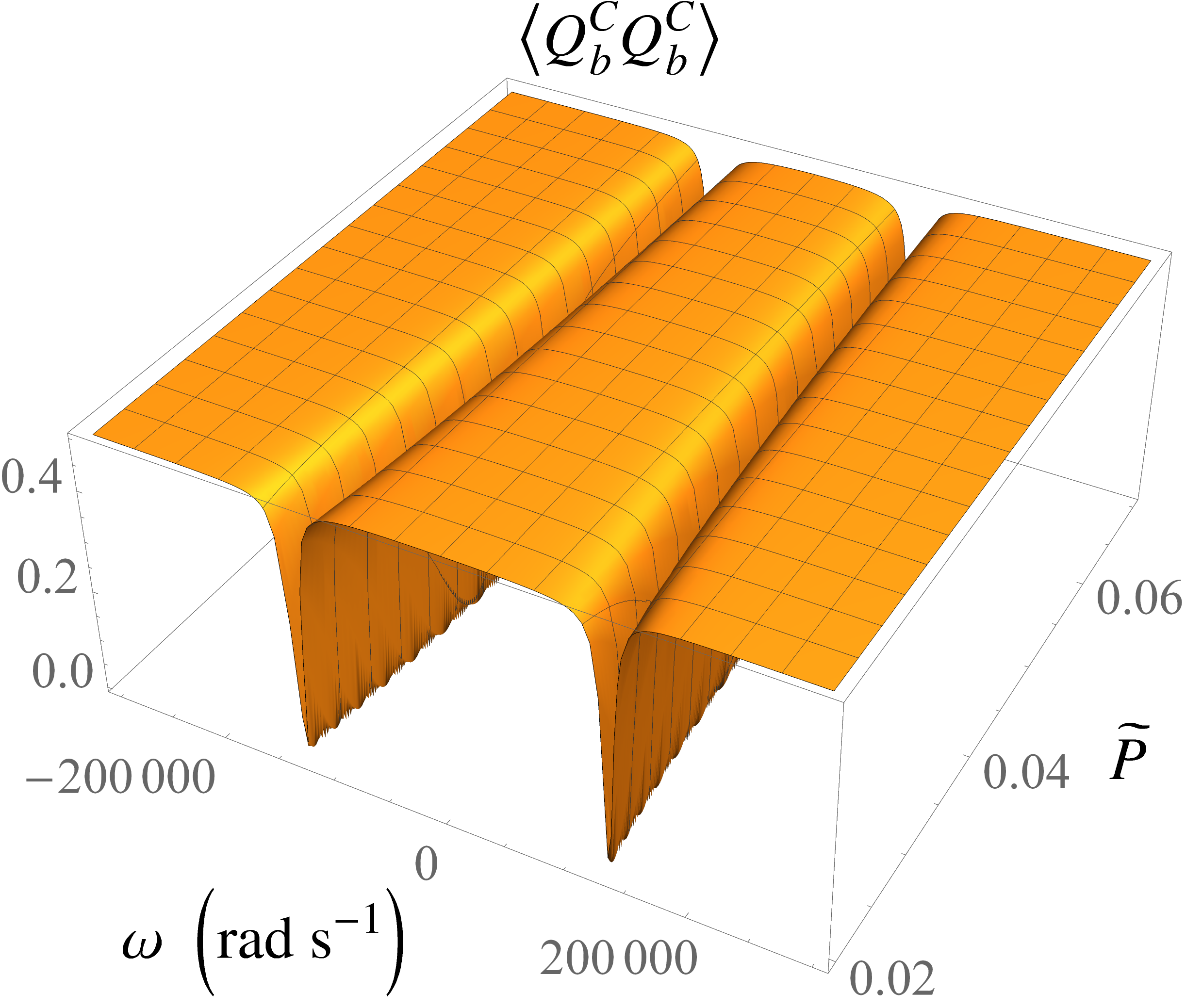}
			\caption{3D plot of the variance of $\widehat{Q}^{C/S}_b$. }
		\end{subfigure}
		\begin{subfigure}[l]{0.8\columnwidth}
			\includegraphics[width=\columnwidth]{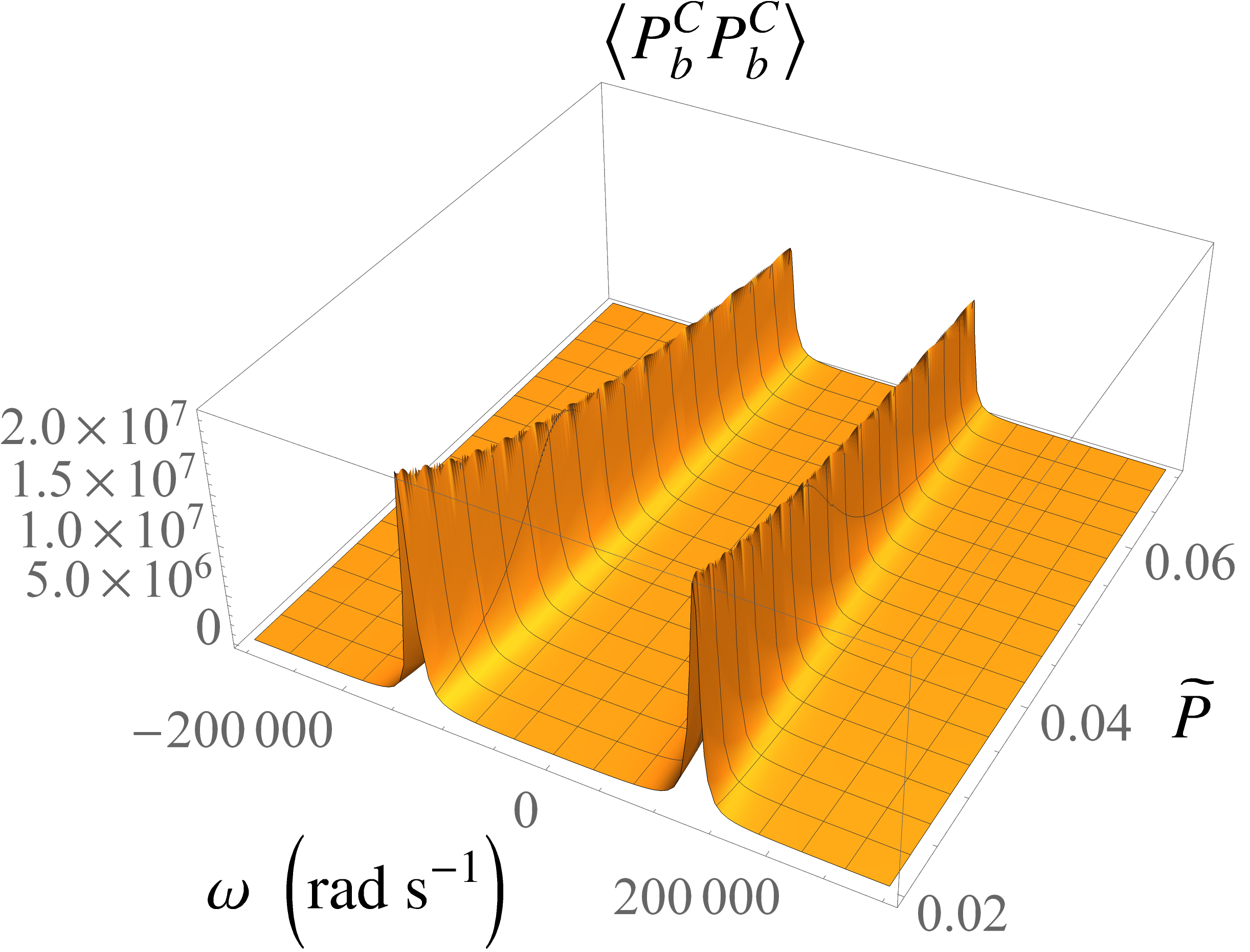}
			\caption{3D plot of the variance of $\widehat{P}^{C/S}_b$.}
		\end{subfigure}
		\caption{3D plots of the variance of cavity and mirror output channels $\widehat{Q}_{a/b}$ and $\widehat{P}_{a/b}$ quadratures. Squeezing can be seen in the mirror output position variance. Parameters as in Fig.\ref{fig:entanglement}}
		\label{fig:mirrorvariance}
	\end{figure*}
	\begin{figure}[h]
	\centering
	\begin{subfigure}[l]{0.8\columnwidth}
			\includegraphics[width=\columnwidth]{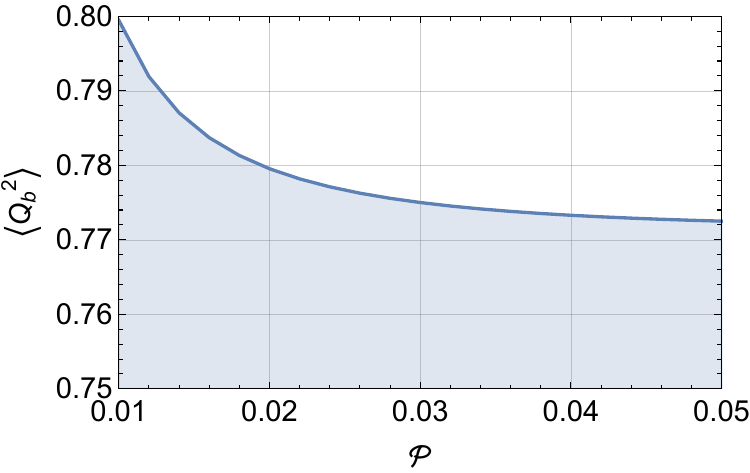}
			\caption{3D plot of the variance of $\widehat{Q}_b$. }
		\end{subfigure}
\par\medskip
	\begin{subfigure}[l]{0.8\columnwidth}
			\includegraphics[width=\columnwidth]{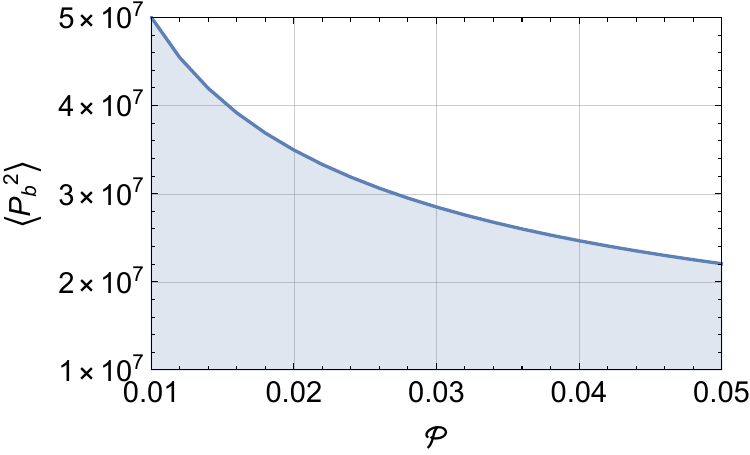}
			\caption{3D plot of the variance of $\widehat{P}_b$.}
		\end{subfigure}
	\caption{Plots of the internal mirror $\widehat{Q}$ and $\widehat{P}$ quadrature variance for the linearised mirror operators $\delta b^{(\dagger)}$. Note that these hold only for a specific phase, chosen here such that both $\hat{P}$ and $\hat{Q}$ values are real, as such, the explicit variance will rotate in phase space, swapping the quadrature variances. No squeezing can be seen in the position or momentum variance. Parameters as in Fig.\ref{fig:entanglement}}
	\label{fig:internalmirrorvariance}
\end{figure}
	\section{Entropy of entanglement \label{sec:entropy}}
	For a bipartite pure Gaussian state, the R\'{e}nyi-2 entropy of entanglement (Von Neumann entropy) is defined as \cite{adesso_continuous_2014}
	\begin{align}
	\mathcal{E}_2\rbracket{\sigma_{a:b}} &= \frac{1}{2}\log_2\sbracket{\text{det}\rbracket{2\sigma_a}}\\
	&=\frac{1}{2}\log_2\sbracket{\text{det}\rbracket{2\sigma_b}}.
	\end{align}
	This characterises the entanglement between the two subsystems, $a$ and $b$. Given an input vacuum state, we can calculate the entanglement between the output of the mirror and cavity (see \cref{fig:entanglement}). At larger $\widetilde{P}$ the coupling strength decreases; this is manifest in the decreasing entanglement at larger $\widetilde{P}$. The location of the peak entanglement is affected by system dynamics. By considering the real part of the determinant of the matrix $T$ and keeping only the leading order terms, we were able to find an approximate expression for the resonant frequency at which the entanglement peaks as $\omega\sim\sqrt{\Omega_Mg_c^2/\left(-\Delta\right)}$, see the appendix for more information.
	 While we have strong entanglement between the mirror and the cavity, our simple theory does not provide for an easy way to access the mirror output. This could be remedied through additional interactions with the mirror. Earlier we suggested that a second cavity could be used to enhance mirror damping at certain frequencies. In the context of a mirror pendulum, such a system has been shown to transfer the entanglement between the cavity and mirror to entanglement between two cavities \cite{PhysRevA.92.022301}. We also note the result of \citeauthor{vanner_cooling-by-measurement_2013} where mechanical state tomography has been shown to `cool-by-measurement' \cite{vanner_cooling-by-measurement_2013}. Thus, accessing the mechanical state through additional interaction with the mirror could be useful in both controlling mirror damping and transfer of entanglement.
	
	In the absence of such interactions we expect the $\hat{b}$ mode to couple to an internal reservoir of phonon modes in the mirror - a point we will take up shortly.

	\section{Quadrature variances \label{sec:quadrature}}
	Now let us consider the quadrature variances of $a$ and $b$ output channels. We do not need to consider the cosine and sine quadratures separately, as the variance of the cosine quadratures are the same as the sine quadratures. As in the case of the entropy of entanglement plot, the central locations of the peaks in \cref{fig:mirrorvariance} are determined by the system dynamics. In \cref{fig:mirrorvariance}, we can see that the variance of $\widehat{Q}_b$ dips below $\frac{1}{2}$, which indicates squeezing. The maximum squeezing is seen at some linear combination of the cosine and sine quadratures and given by the eigenvalues of $\sigma_b$. However, for our particular parameters, the detuning is small so while there is some coupling between the cosine and sine quadratures, there is less than $0.1\%$ difference between our plots and the maximum squeezing.  However, this squeezing is not seen in the cavity field output or indeed even in the internal mirror quadrature (as opposed the mirror output's quadrature) as shown in \cref{fig:internalmirrorvariance}. Here the internal mirror perturbation quadrature is found as the integral over all positive frequency modes without any sine or cosine mix-down, that is
	\begin{align}
		\left\langle \hat{Q}_b^2\right\rangle=&\int_{0}^{\infty} \delta\hat{b}^{\dagger}(\omega)+\delta\hat{b}(\omega)~\textrm{d}\omega\\
		\left\langle \hat{P}_b^2\right\rangle=&\int_{0}^{\infty} \delta\hat{b}^{\dagger}(\omega)-\delta\hat{b}(\omega)~\textrm{d}\omega
	\end{align}
	Note that this differs by the appropriate scale factors from the actual position and momentum quadrature as it corresponds to the linearised perturbation operators ($\delta b^{(\dagger)}$) as given in equations \ref{eq:position perturbation} and \ref{eq:momentum perturbation}. Given the squeezing appears only in the mirror output it can be understood as the mirror uncertainty being driven primarily by the laser field. The entanglement visible in the entropy will appear in this output as a spectral response, however, this will then rapidly thermalise as the energy is shared among the mirrors internal phononic modes. To that end, we now consider the thermal effects of the mirror.
	
	\section{Mirror Temperature \label{sec:temperature}}
	We now consider the temperature of the mirror due to the driven mechanical oscillation and absorption. In order to assign a temperature to these effects we must first make a few assumptions about the mirror. We will consider the fluctuation energy of the mirror to be transferred into other mechanical modes of the mirror with a thermal distribution. This requires that these mechanical modes can be treated as a reservoir, indeed as a Markovian bath. We also assume that initially, the temperature of these modes is zero to ensure we are only modelling the induced temperature. Finally, we will assume the mirror to be a cylindrical disk of thickness $\tau=100~\mu$m with a density $\rho\approx2\times10^3$ kgm$^{-3}$, similar to silicon. Fixing $\tau$ and $\rho$ is necessary to ensure the surface area and mass scale proportionally so we can still work with the scaled power $\tilde{P}$.
		
\begin{figure}[h]
			\centering
			\includegraphics[width=0.9\columnwidth]{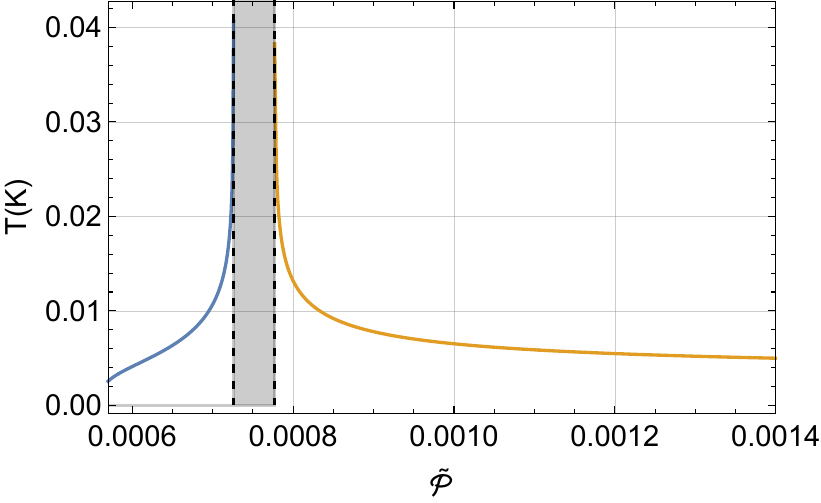}
			\caption{A plot of the mechanical osciallator temperature in the two separated stable regions (blue and orange) as a function of dimensionless laser power, $\widetilde{P}$. Note that when $\widetilde{P}\approx0.00056$ and $0.00072\le\widetilde{P}\le0.00078$, in the greyed out region, the mirror is unstable. Parameters are: $L=\SI{5}{\centi\meter}$, $\lambda_L= \SI{1050}{\nano\metre}$, $\kappa = \SI{1.35e7}{\radian\per\second}$ and $\Gamma=\SI{1e4}{\radian\per\second}$. A lower power range is shown here to reveal the unusual temperature scaling behaviour.}
			\label{fig:Mechanical-Temp}
		\end{figure}

		To model the mechanical fluctuation induced temperature we calculate the energy flux, scaled by $\left(mg\right)^{-1}$, due to the phonon output:
		\begin{equation}
			\tilde{E}_{\textrm{flux}}=\int_{0}^{\infty}\hbar\omega\left\langle \hat{b}^{\dagger}_{\textrm{out}}(\omega)\hat{b}_{\textrm{out}}(\omega)\right\rangle.
		\end{equation}
		The temperature can then be related to this using the Stefan-Boltzmann law to give
		\begin{equation}
			T=\left(\frac{\tilde{E}_{\textrm{flux}}g \rho \tau}{2\sigma}\right)^{1/4}.
		\end{equation}
		where $\sigma=\frac{2\pi^5k_B^4}{15c^2h^3}=5.67\times10^{-8}$Wm$^{-2}$K$^{-4}$. The resulting temperature is shown in Figure~\ref{fig:Mechanical-Temp}. Interestingly, this predicts two separate regions with opposite scaling. In the low power region, the induced temperature increases with input power in a manner that is perhaps unsurprising. The higher power scaling, which covers most of the stable parameter space, shows the temperature actually decreasing with increasing power. The intermediate region corresponds to unstable solutions with a corresponding apparent spike in temperature.
		
		In principle, black-body radiation from the mirror will retain the entanglement with the laser, although the frequency dependence will be scrambled.
		
\begin{figure}[h]
			\centering
			\includegraphics[width=0.9\columnwidth]{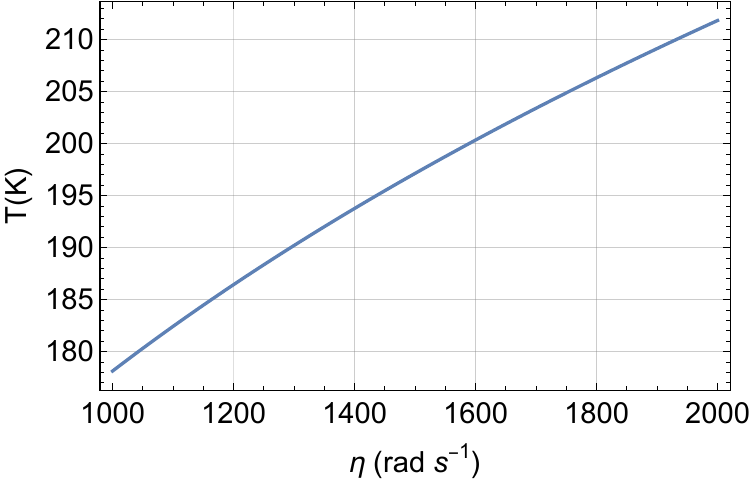}
			\caption{A plot of the temperature of the mirror due to absorption of the cavity field as a function of the coupling rate $\widetilde{\eta}$. Note that when $\eta\approx1500$ rad s$^{-1}$ corresponds to a reflectivity of $0.999999$.}
		\label{fig:Absorption-Temp}
	\end{figure}

		Up to now we have neglected the effect of absorption of the coherent laser field by the mirror. To include this we allow $\eta\ne0$ in Equation \ref{eq:adot} and add a third output channel $\hat{a}_{m, out}$ where $\hat{a}_{m, out}=\hat{a}_{m, in}-\sqrt{\eta}\hat{a}$. This channel corresponds to absorption of the cavity field by the mirror. While this does change the steady state solutions, it does not have a significant impact other than requiring a larger input power to balance the added loss. Preceding similarly to above, and adjusting the input power such that the steady state equations match the values used throughout this manuscript, we find that the absorption induced temperature is far higher than that due to the mirror quantum fluctuations. Figure \ref{fig:Absorption-Temp} shows the mirror steady state temperature as a function of the coupling rate $\eta$. The coupling rate is related to the mirror reflectance $r$ by
		\begin{equation}
			\eta=\frac{c}{2L}\left(1-r\right)
		\end{equation}
		so $\eta=1500$ rad Hz corresponds to a mirror that is $99.9999\%$ reflective. Therefore, any absorption of the cavity field by the mirror will likely swamp a thermal signal due to the mirror oscillation, and hence any observable entanglement with the laser.

	\section{Conclusion \label{sec:conclusion}}
	We have proposed a one dimensional optomechanical system where an upper mirror is levitated and supported alone by the radiation force from a cavity. While the system has two steady-state solutions---a blue and a red detuned steady state---the mirror is only stable for the blue detuned steady-state with sufficient damping on the mirror and cavity. The unavoidable detuning leads to coupling between the symmetric and antisymmetric quadrature side-bands leading to the introduction of an eight dimensional covariance matrix. With a vacuum input state, we find entanglement between the output of the mirror and cavity and squeezing in the output position of the mirror. This squeezed spectral response is expected to immediately thermalise. In principle strong entanglement is predicted between the laser and mirror radiation, however, this will be swamped by absorption heating for reasonable parameters. Avoiding this issue would require a different method for extracting energy from the mirror.
	\\
	\section*{Acknowledgements}
	C.T.M.H. would like to thank Jinyong Ma for invaluable discussions during a visit to the Australian National University and Ping Koy Lam for financial support during said visit. TCR acknowledges useful discussions with Warwick Bowen. This research was supported by the Australian Government Research Training Program (RTP) Scholarship, the Natural Sciences and Engineering Research Council of Canada and the Australian Research Council Centre of Excellence for Quantum Computation and Communication Technology (Project No. CE170100012)

\begin{widetext}	
\appendix

\section{Steady-state solution}
With our steady-state conditions, we find that the steady-state equations are,
\begin{align}
	&\sbracket{\frac{\kappa^2}{4} + \rbracket{\Omega_c ( q_\text{SS})-\Omega_L}^2}N_c = \kappa N_\text{in} \label{eq:steadystatealpha}\\
	&p_\text{SS} = 0 \\
	&0= mg - \frac{\hbar \Omega_c^2(q_\text{SS})}{j\pi c} N_c 
\end{align}
where $\mathcal{O}_\text{SS} = \Braket{\mathcal{O}}$. The two steady state solutions are,
\begin{align}
	q_1 &= \frac{gm\sbracket{-4 j c \pi \Omega_L + L\rbracket{\kappa^2+4\Omega_L^2}} - 2 \sqrt{\pi } \sqrt{ c g m j  \sbracket{-c g j m \pi \kappa^2 + N_\text{in} \kappa \hbar \rbracket{\kappa^2 + 4 \Omega_L^2}}}}{g m \rbracket{\kappa^2 + 4 \Omega_L^2}},\\
	N_{c,1} &= \frac{-4cgm j\pi \rbracket{\kappa^2-4 \Omega_L^2} + 4 \kappa N_\text{in}\hbar \rbracket{\kappa^2+4\Omega_L^2} + 16 \sqrt{\pi} \Omega_L \sqrt{ c g mj  \sbracket{-c g m j\pi \kappa^2 + N_\text{in} \kappa \hbar \rbracket{\kappa^2 + 4 \Omega_L^2}}}  }{\hbar \rbracket{\kappa^2 + 4 \Omega_L^2}^2}
\end{align}
and
\begin{align}
	q_2 &= \frac{gm\sbracket{-4 jc \pi \Omega_L + L\rbracket{\kappa^2+4\Omega_L^2}} +2 \sqrt{\pi } \sqrt{ c g mj  \sbracket{-c g jm \pi \kappa^2 + N_\text{in} \kappa \hbar \rbracket{\kappa^2 + 4 \Omega_L^2}}}}{g m \rbracket{\kappa^2 + 4 \Omega_L^2}},\\
	N_{c,2} &= \frac{-4cgmj\pi \rbracket{\kappa^2-4 \Omega_L^2} + 4 \kappa N_\text{in}\hbar \rbracket{\kappa^2+4\Omega_L^2} - 16 \sqrt{\pi} \Omega_L \sqrt{ c g mj  \sbracket{-c g mj \pi \kappa^2 + N_\text{in} \kappa \hbar \rbracket{\kappa^2 + 4 \Omega_L^2}}}  }{\hbar\rbracket{\kappa^2 + 4 \Omega_L^2}^2}.
\end{align}
\subsection{Detuning of the steady state solutions}
We can express the detuning $\Delta = \Omega_c - \Omega_L$ in terms of the steady state solutions,
\begin{align}
	\Delta_1 =& \frac{2N_\text{in} \kappa\Omega_L \hbar -\sqrt{\pi } \sqrt{ c g m j  \sbracket{-c g j m \pi \kappa^2 + N_\text{in} \kappa \hbar \rbracket{\kappa^2 + 4 \Omega_L^2}}}}{2cgmj\pi - 2 N_\text{in} \kappa \hbar}\\
	\Delta_2 & =  \frac{2N_\text{in} \kappa\Omega_L \hbar +\sqrt{\pi } \sqrt{ c g m j  \sbracket{-c g j m \pi \kappa^2 + N_\text{in} \kappa \hbar \rbracket{\kappa^2 + 4 \Omega_L^2}}}}{2cgmj\pi - 2 N_\text{in} \kappa \hbar}.
\end{align}
To determine which solution is blue or red detuned we need reasonable estimates of the system parameters. As was argued in the main text we will need $\widetilde{P}= \frac{\hbar N_\text{in} \Omega_L}{mgc}$. Together with $j =\text{Round}\rbracket{\frac{L\Omega_L}{\pi c}}$, we can see that the parameter dependence of the detuning is $\Delta = \Delta ( \widetilde{P}, \Omega_L,L,\kappa)$. Thus, the denominator is approximately $2cgm \rbracket{j\pi -\frac{\kappa}{\Omega_L}}$; since we require that  $\frac{\kappa}{\Omega_L}\ll 1$, we know for any good cavity, the denominator is always positive. This means that only the first solution can be blue detuned ($\Delta<0$). Using the same argument that we must have a good cavity, we can approximate the numerator as
\begin{align}
	&2N_\text{in} \kappa\Omega_L \hbar -\sqrt{\pi } \sqrt{ c g m j  \sbracket{-c g j m \pi \kappa^2 + N_\text{in} \kappa \hbar \rbracket{\kappa^2 + 4 \Omega_L^2}}}\nonumber\\
	&\approx 2N_\text{in} \kappa\Omega_L \hbar -\sqrt{\pi } \sqrt{ c g m j  N_\text{in} \kappa \hbar  4 \Omega_L^2}.
\end{align}
Substituting our rough estimate of $N_\text{in}$, we find that
\begin{align}
	&2N_\text{in} \kappa\Omega_L \hbar -\sqrt{\pi } \sqrt{ c g m j  N_\text{in} \kappa \hbar  4 \Omega_L^2}\nonumber\\
	&=2\kappa mgc -  mgc\sqrt{\pi j\kappa \Omega_L}
\end{align}
and finally, we know that $j>0, j \in \mathbb{N}$ so $\sqrt{\pi j}>1$ and again, $\frac{\kappa}{\Omega_L}\ll 1$, so $\sqrt{\Omega_L\kappa}\gg \kappa$. This ensures that for a good cavity we must have $\Delta_1<0$.

We thus conclude that we have two steady state solutions, one above the resonance point (blue detuned where $\Delta <0$) and one below (red detuned where $\Delta >0$).

\section{Quadratures}

The quadratures and entropy are based of the perturbation operator's equations of motion in frequency space given by
\begin{equation}\label{Tsol}
	\rbracket{\hat{b},\hat{b}^\dagger,\hat{a},\hat{a}^\dagger} = T \rbracket{\sqrt{\Gamma} \hat{b}_\text{in},\sqrt{\Gamma} \hat{b}_\text{in}^\dagger,\sqrt{\kappa} \hat{a}_\text{in},\sqrt{\kappa} \hat{a}_\text{in}^\dagger}
\end{equation}
where $T= \rbracket{-i \omega I -A}^{-1}$.

From the input-output relations, we have,
\begin{align}
	a_\text{out}(\omega) &= a_\text{in}(\omega)-\sqrt{\kappa}  a(\omega) \\
	b_\text{out}(\omega) &= b_\text{in}(\omega)-\sqrt{\Gamma}  b(\omega) 
\end{align}
From \eqref{Tsol}  
we have 
\begin{align}
	a &= T_{31}\sqrt{\Gamma} b_\text{in} + T_{32}\sqrt{\Gamma} b_\text{in}^\dagger + T_{33} \sqrt{\kappa} a_\text{in} + T_{34}\sqrt{\kappa} a_\text{in}^\dagger\\
	a^\dagger &=T_{41}\sqrt{\Gamma} b_\text{in} + T_{42}\sqrt{\Gamma} b_\text{in}^\dagger + T_{43} \sqrt{\kappa} a_\text{in} + T_{44}\sqrt{\kappa} a_\text{in}^\dagger\\
	b &= T_{11}\sqrt{\Gamma} b_\text{in} + T_{12}\sqrt{\Gamma} b_\text{in}^\dagger + T_{13} \sqrt{\kappa} a_\text{in} + T_{14}\sqrt{\kappa} a_\text{in}^\dagger\\
	b^\dagger &=T_{21}\sqrt{\Gamma} b_\text{in} + T_{22}\sqrt{\Gamma} b_\text{in}^\dagger + T_{23} \sqrt{\kappa} a_\text{in} + T_{24}\sqrt{\kappa} a_\text{in}^\dagger
\end{align}
and our quadrature definitions are $X^\pm_a = \frac{1}{\sqrt{2}}\rbracket{a\pm a^\dagger}$, $X^\pm_b = \frac{1}{\sqrt{2}}\rbracket{b\pm b^\dagger}$, which give us the quadratures for the cavity
\begin{align}
	X_{a_\text{out}}^\pm &= \frac{1}{\sqrt{2}} \cbracket{-\rbracket{T_{31}\pm T_{41}} \sqrt{\kappa\Gamma} b_\text{in}-\rbracket{T_{32}\pm T_{42}} \sqrt{\kappa\Gamma} b_\text{in}^\dagger-\sbracket{\kappa\rbracket{T_{33}\pm T_{43}} -1} a_\text{in}-\sbracket{\kappa\rbracket{T_{34}\pm T_{44}} \mp1}a_\text{in}^\dagger},
\end{align}
and  for the mirror
\begin{align}
	X_{b_\text{out}}^\pm &=\frac{1}{\sqrt{2}}\cbracket{-\sbracket{\Gamma\rbracket{T_{11}\pm T_{21}} -1} b_\text{in}-\sbracket{\Gamma\rbracket{T_{12}\pm T_{22}}\mp 1} b_\text{in}^\dagger-\rbracket{T_{13}\pm T_{23}}\sqrt{\kappa\Gamma} a_\text{in}-\rbracket{T_{14}\pm T_{24}} \sqrt{\kappa\Gamma}a_\text{in}^\dagger}.
\end{align}
Let us change notation to keep track of the frequency dependence of our operators. Note that because $X^\pm(\omega) = a(\omega) \pm a(-\omega)^\dagger = \pm X^\pm(-\omega)^\dagger$. This motivates us to define the following functions,
\begin{align}
	B^\pm_{a_\text{out}}(\omega) &=  -\rbracket{T_{31}(\omega)\pm T_{41}(\omega)} \sqrt{\kappa\Gamma}, \\
	\pm B^\pm_{a_\text{out}}(-\omega)^*&=-\rbracket{T_{32}(\omega)\pm T_{42}(\omega)} \sqrt{\kappa\Gamma} ,\\
	A^\pm_{a_\text{out}}(\omega)&=-\sbracket{\kappa\rbracket{T_{33}(\omega)\pm T_{43}(\omega)} -1},\\
	\pm A^\pm_{a_\text{out}}(-\omega)^* & = -\sbracket{\kappa\rbracket{T_{34}(\omega)\pm T_{44}(\omega)} \mp1},\\
	B^\pm_{b_\text{out}}(\omega) &=  -\sbracket{\Gamma\rbracket{T_{11}\pm T_{21}} -1}, \\
	\pm B^\pm_{b_\text{out}}(-\omega)^*&=-\sbracket{\Gamma\rbracket{T_{12}\pm T_{22}}\mp 1},\\
	A^\pm_{b_\text{out}}(\omega)&=-\sqrt{\kappa\Gamma}\rbracket{T_{13}(\omega)\pm T_{23}(\omega)}, \\
	\pm A^\pm_{b_\text{out}}(-\omega)^* & = -\sqrt{\kappa\Gamma}\rbracket{T_{14}(\omega)\pm T_{24}(\omega)}.
\end{align}
The quadratures may now be defined as
\begin{subequations}
	\begin{align}
		X_{a_\text{out}}^\pm &= \frac{1}{\sqrt{2}}\cbracket{B^\pm_{a_\text{out}}(\omega)  b_\text{in}(\omega)\pm B^\pm_{a_\text{out}}(-\omega)^* b_\text{in}(-\omega)^\dagger +A^\pm_{a_\text{out}}(\omega) a_\text{in}(\omega)\pm A^\pm_{a_\text{out}}(-\omega)^* a_\text{in}(-\omega)^\dagger},\\
		X_{b_\text{out}}^\pm &=\frac{1}{\sqrt{2}}\cbracket{ B^\pm_{b_\text{out}}(\omega)  b_\text{in}(\omega)\pm B^\pm_{b_\text{out}}(-\omega)^* b_\text{in}(-\omega)^\dagger +A^\pm_{b_\text{out}}(\omega) a_\text{in}(\omega)\pm A^\pm_{b_\text{out}}(-\omega)^* a_\text{in}(-\omega)^\dagger}.
	\end{align}
	\label{eq:quadratures}
\end{subequations}

Let us introduce the notation $X^\alpha_{i_\text{out}}$, where $\alpha, \beta, \ldots \in \cbracket{+,-}$ indicate $+$ or $-$ and let $i, j,\ldots \in \cbracket{a,b}$ indicate the mode. Then we can combine the above equations into
\begin{equation}
	X_{i_\text{out}}^\alpha = \frac{1}{\sqrt{2}}\cbracket{B^\alpha_{i_\text{out}}(\omega)  b_\text{in}(\omega)+\alpha B^\alpha_{i_\text{out}}(-\omega)^* b_\text{in}(-\omega)^\dagger +A^\alpha_{i_\text{out}}(\omega) a_\text{in}(\omega)+\alpha A^\alpha_{i_\text{out}}(-\omega)^* a_\text{in}(-\omega)^\dagger}.
\end{equation}

\section{Covariance Matrix}

In this section we   list the matrix elements of the covariance matrix. 

It will first be useful to  examine the submatrices. The submatrix for $b$ is
\begin{align}
	\sigma_b =  \Braket{0|\begin{pmatrix}
			Q_b^{C} Q_b^{C} &\frac{1}{2} \rbracket{Q_b^{C}  P_b^{C} +P_b^{C} Q_b^{C}}& \frac{1}{2}\rbracket{Q_b^{C}   Q_b^{S} + Q_b^{S} Q_b^{C}}&\frac{1}{2} \rbracket{Q_b^{C} P_b^{S}  + P_b^{S} Q_b^{C} }\\
			& P_b^{C} P_b^{C}  &\frac{1}{2} \rbracket{ P_b^{C} Q_b^{S} + Q_b^{S} P_b^{C} } &\frac{1}{2} \rbracket{ P_b^{C}   P_b^{S} + P_b^{S}  P_b^{C}}\\
			&   &Q_b^{S}  Q_b^{S} & \frac{1}{2} \rbracket{Q_b^{S}  P_b^{S} + P_b^{S}  Q_b^{S}}\\
			&  &  & P_b^{S} P_b^{S} 
		\end{pmatrix}|0},
\end{align}
while the submatrix for $a$ is,
\begin{align}
	\sigma_a =  \Braket{0|\begin{pmatrix}
			Q_a^{C} Q_a^{C} &\frac{1}{2} \rbracket{Q_a^{C}  P_a^{C} +P_a^{C} Q_a^{C}}& \frac{1}{2}\rbracket{Q_a^{C}   Q_a^{S} + Q_a^{S} Q_a^{C}}&\frac{1}{2} \rbracket{Q_a^{C} P_a^{S}  + P_a^{S} Q_a^{C} }\\
			& P_a^{C} P_a^{C}  &\frac{1}{2} \rbracket{ P_a^{C} Q_a^{S} + Q_a^{S} P_a^{C} } &\frac{1}{2} \rbracket{ P_a^{C}   P_a^{S} + P_a^{S}  P_a^{C}}\\
			&   &Q_a^{S}  Q_a^{S} & \frac{1}{2} \rbracket{Q_a^{S}  P_a^{S} + P_a^{S}  Q_a^{S}}\\
			&  &  & P_a^{S} P_a^{S}  
		\end{pmatrix}|0}.
\end{align}
The upper right matrix is
\begin{align}
	\sigma_\text{upper} =  \frac{1}{2}\Braket{0|\begin{pmatrix}
			{Q_b^{C}   Q_a^{C} + Q_a^{C} Q_b^{C}}& {Q_b^{C} P_a^{C}  + P_a^{C} Q_b^{C} }& {Q_b^{C}   Q_a^{S} + Q_a^{S} Q_b^{C}}& {Q_b^{C} P_a^{S}  + P_a^{S} Q_b^{C} }\\
			{P_b^{C}   Q_a^{C} + Q_a^{C} P_b^{C}}	&  {P_b^{C} P_a^{C}  + P_a^{C} P_b^{C} }  &{P_b^{C}   Q_a^{S} + Q_a^{S} P_b^{C}} & {P_b^{C} P_a^{S}  + P_a^{S} P_b^{C} }\\
			{Q_b^{S}   Q_a^{C} + Q_a^{C} Q_b^{S}}	&  {Q_b^{S} P_a^{C}  + P_a^{C} Q_b^{S} }  &{Q_b^{S}   Q_a^{S} + Q_a^{S} Q_b^{S}} &  {Q_b^{S} P_a^{S}  + P_a^{S} Q_b^{S} }\\
			{P_b^{S}   Q_a^{C} + Q_a^{C} P_b^{S}}	&  {P_b^{S} P_a^{C}  + P_a^{C} P_b^{S} } & {P_b^{S}   Q_a^{S} + Q_a^{S} P_b^{S}} &  {P_b^{S} P_a^{S}  + P_a^{S} P_b^{S} }  
		\end{pmatrix}|0}.
\end{align}
To calculate these matrix elements, it is useful to introduce an intermediate expression for the cosine
\begin{align}
	X^{\alpha,C}_i(\omega) &= \sqrt{\frac{2}{\pi}} \int \diff{t} \frac{\cos\rbracket{\omega t}}{\sqrt{2}} X^\alpha_i(t) \nonumber\\
	&=\frac{1}{\sqrt{2}}\rbracket{ X^\alpha_i(\omega) + X^\alpha_i(-\omega)},
\end{align}
and sine quadrature,
\begin{align}
	X^{\alpha,S}_i(\omega) &= \sqrt{\frac{2}{\pi}} \int \diff{t}\frac{\sin\rbracket{\omega t}}{\sqrt{2}} X^\alpha_i(t) \nonumber\\
	&=\frac{-i}{\sqrt{2}} \rbracket{X^\alpha_i(\omega) - X^\alpha_i(-\omega)}.
\end{align}
Which can be combined into one expression,
\begin{align}
	X^{\alpha,\aleph}_i (\omega) = \frac{1}{\sqrt{2}}\rbracket{\aleph X^\alpha_i(\omega) + X^\alpha_i(-\omega)},
\end{align}
where $\aleph = \pm 1$, $X^{\alpha,C}_i = X^{\alpha,+}_i$ and $X^{\alpha,S}_i = iX^{\alpha,-}_i$. These sine and cosine mixdowns are only physical if they are integrated over a finite time with the appropriate normalised mode function. Implementing this will promote the operators to mode operators which are sharply, but not infinitely, peaked at a specific frequency $\omega$. As such, the commutator become $\left[a(\omega),a(\omega')^{\dagger}\right]=\delta_{\omega,\omega'}$. The conversion to the quadratures used in the main text are as follows.
\begin{align}
	Q^C_i(\omega) &=  \frac{1}{\sqrt{2}}\rbracket{ X^+_i(\omega) + X^+_i(-\omega)},\\
	P^C_i(\omega) &=  \frac{1}{\sqrt{2}} \rbracket{-i X^-_i(\omega) +(-i) X^-_i(-\omega)}\nonumber\\
	&= \frac{-i}{\sqrt{2}} \rbracket{ X^-_i(\omega) + X^-_i(-\omega)},\\
	Q^S_i(\omega) &=  \frac{-i}{\sqrt{2}}\rbracket{ X^+_i(\omega) - X^+_i(-\omega)},\\
	P^S_i(\omega) &=  \frac{-i}{\sqrt{2}} \rbracket{-i X^-_i(\omega) -(-i) X^-_i(-\omega)}\nonumber\\
	&= \frac{-1}{\sqrt{2}} \rbracket{ X^-_i(\omega) - X^-_i(-\omega)}.
\end{align}
With this definition the matrix elements are---modulo factors of $-i$---composed of,
\begin{align}
	&\frac{1}{2}\Braket{X^{\alpha,\aleph}_{i_\text{out}}(\omega) X^{\beta,\beth}_{j _\text{out}}(\omega)+ X^{\beta,\beth}_{j_\text{out}} (\omega)  X^{\alpha,\aleph}_{i_\text{out}}( \omega) } \\
	&= \frac{1}{2}\begin{cases}
		\beth \frac{1}{2}\Braket{X^\alpha_{i_\text{out}}(\omega) X^\beta_{j _\text{out}}(-\omega)+ X^\beta_{j_\text{out}} (\omega)  X^\alpha_{i_\text{out}}(- \omega) } + \beth \frac{1}{2}\Braket{X^\alpha_{i_\text{out}}(-\omega) X^\beta_{j _\text{out}}(\omega)+ X^\beta_{j_\text{out}} (-\omega)  X^\alpha_{i_\text{out}}(\omega) }, &\aleph = \beth\\
		\beth \frac{1}{2}\Braket{-X^\alpha_{i_\text{out}}(\omega) X^\beta_{j _\text{out}}(-\omega)+ X^\beta_{j_\text{out}} (\omega)  X^\alpha_{i_\text{out}}(- \omega) } + \beth \frac{1}{2}\Braket{X^\alpha_{i_\text{out}}(-\omega) X^\beta_{j _\text{out}}(\omega)- X^\beta_{j_\text{out}} (-\omega)  X^\alpha_{i_\text{out}}(\omega) }, &\aleph = -\beth
	\end{cases}.
\end{align}
We can now calculate the expectation, where we find that 
\begin{equation}
	\frac{1}{2}\Braket{X^\alpha_{i_\text{out}}(\omega) X^\beta_{j _\text{out}}(-\omega)+ X^\beta_{j_\text{out}} (\omega)  X^\alpha_{i_\text{out}}(- \omega) } = \frac{1}{2}\begin{cases}
		\beta \Re{\sbracket{ B_{i_\text{out}}^\alpha (\omega) B_{j_\text{out}}^\beta (\omega)^*+ A_{i_\text{out}}^\alpha (\omega) A_{j_\text{out}}^\beta (\omega)^*}}, & \alpha = \beta\\
		i \beta \Im{\sbracket{ B_{i_\text{out}}^\alpha (\omega) B_{j_\text{out}}^\beta (\omega)^*+ A_{i_\text{out}}^\alpha (\omega) A_{j_\text{out}}^\beta (\omega)^*}}, & \alpha = -\beta
	\end{cases},
\end{equation}
\begin{equation}
	\frac{1}{2}\Braket{X^\alpha_{i_\text{out}}(\omega) X^\beta_{j _\text{out}}(-\omega)- X^\beta_{j_\text{out}} (\omega)  X^\alpha_{i_\text{out}}(- \omega) } = \frac{1}{2}\begin{cases}
		i\beta \Im{\sbracket{ B_{i_\text{out}}^\alpha (\omega) B_{j_\text{out}}^\beta (\omega)^*+ A_{i_\text{out}}^\alpha (\omega) A_{j_\text{out}}^\beta (\omega)^*}}, & \alpha = \beta\\
		\beta \Re{\sbracket{ B_{i_\text{out}}^\alpha (\omega) B_{j_\text{out}}^\beta (\omega)^*+ A_{i_\text{out}}^\alpha (\omega) A_{j_\text{out}}^\beta (\omega)^*}}, & \alpha = -\beta
	\end{cases}.
\end{equation}
This then gives us,
\begin{align}
	&\frac{1}{2}\Braket{X^{\alpha,\aleph}_{i_\text{out}}(\omega) X^{\beta,\beth}_{j _\text{out}}(\omega)+ X^{\beta,\beth}_{j_\text{out}} (\omega)  X^{\alpha,\aleph}_{i_\text{out}}( \omega) }\nonumber \\
	&=\frac{1}{4}\begin{cases}
		\cbracket{\beta \beth \Re{\sbracket{ B_{i_\text{out}}^\alpha (\omega) B_{j_\text{out}}^\beta (\omega)^*+ A_{i_\text{out}}^\alpha (\omega) A_{j_\text{out}}^\beta (\omega)^*}} + \rbracket{\omega \rightarrow -\omega}}, & \alpha = \beta, \aleph = \beth\\
		\cbracket{i \beta \beth \Im{\sbracket{ B_{i_\text{out}}^\alpha (\omega) B_{j_\text{out}}^\beta (\omega)^*+ A_{i_\text{out}}^\alpha (\omega) A_{j_\text{out}}^\beta (\omega)^*}}+ \rbracket{\omega \rightarrow -\omega}}, & \alpha = -\beta, \aleph = \beth\\
		\cbracket{-i\beta\beth \Im{\sbracket{ B_{i_\text{out}}^\alpha (\omega) B_{j_\text{out}}^\beta (\omega)^*+ A_{i_\text{out}}^\alpha (\omega) A_{j_\text{out}}^\beta (\omega)^*}} -\rbracket{\omega \rightarrow -\omega}}, & \alpha = \beta,\aleph = -\beth\\
		\cbracket{-\beta\beth \Re{\sbracket{ B_{i_\text{out}}^\alpha (\omega) B_{j_\text{out}}^\beta (\omega)^*+ A_{i_\text{out}}^\alpha (\omega) A_{j_\text{out}}^\beta (\omega)^*}}-\rbracket{\omega \rightarrow -\omega}}, & \alpha = -\beta,\aleph=-\beth
	\end{cases}. \label{eq:covarianceexpressions}
\end{align}
$\rbracket{\omega \rightarrow -\omega}$ is taken to mean the previous term with $\omega$ replaced with $-\omega$. If we restore the factors of $i$ in our covariance matrices we find the following expressions for the matrix elements.
\begin{align}
	&\frac{1}{2}\Braket{Q^{C}_{i_\text{out}}(\omega) Q^{C}_{j _\text{out}}(\omega)+ Q^C_{j_\text{out}} (\omega)  Q^C_{i_\text{out}}( \omega) } \nonumber\\
	&=\frac{1}{2}\Braket{Q^{S}_{i_\text{out}}(\omega) Q^{S}_{j _\text{out}}(\omega)+ Q^S_{j_\text{out}} (\omega)  Q^S_{i_\text{out}}( \omega) }\nonumber \\
	&= \frac{1}{4}\cbracket{\Re{\sbracket{ B_{i_\text{out}}^+ (\omega) B_{j_\text{out}}^+ (\omega)^*+ A_{i_\text{out}}^+ (\omega) A_{j_\text{out}}^+ (\omega)^*}} + \rbracket{\omega \rightarrow -\omega}}\delta(0)
\end{align}
\begin{align}
	&\frac{1}{2}\Braket{Q^{C}_{i_\text{out}}(\omega) Q^{S}_{j _\text{out}}(\omega)+ Q^S_{j_\text{out}} (\omega)  Q^C_{i_\text{out}}( \omega) } \nonumber\\
	&= -\frac{1}{2}\Braket{Q^{S}_{i_\text{out}}(\omega) Q^{C}_{j _\text{out}}(\omega)+ Q^C_{j_\text{out}} (\omega)  Q^S_{i_\text{out}}( \omega) }\nonumber\\
	&= \frac{1}{4}\cbracket{-\Im{\sbracket{ B_{i_\text{out}}^+ (\omega) B_{j_\text{out}}^+ (\omega)^*+ A_{i_\text{out}}^+ (\omega) A_{j_\text{out}}^+ (\omega)^*}}- \rbracket{\omega \rightarrow -\omega}}\delta(0) 
\end{align}

\begin{align}
	&\frac{1}{2}\Braket{P^{C}_{i_\text{out}}(\omega) P^{C}_{j _\text{out}}(\omega)+ P^{C}_{j_\text{out}} (\omega)  P^{C}_{i_\text{out}}( \omega) } \nonumber\\
	&=\frac{1}{2}\Braket{P^{S}_{i_\text{out}}(\omega) P^S_{j _\text{out}}(\omega)+ P^S_{j_\text{out}} (\omega)  P^{S}_{i_\text{out}}( \omega) } \nonumber\\
	&= \frac{1}{4}\cbracket{\Re{\sbracket{ B_{i_\text{out}}^- (\omega) B_{j_\text{out}}^- (\omega)^*+ A_{i_\text{out}}^- (\omega) A_{j_\text{out}}^- (\omega)^*}} + \rbracket{\omega \rightarrow -\omega}}\delta(0)
\end{align}

\begin{align}
	&\frac{1}{2}\Braket{P^{C}_{i_\text{out}}(\omega) P^{S}_{j _\text{out}}(\omega)+ P^{S}_{j_\text{out}} (\omega)  P^{C}_{i_\text{out}}( \omega) } \nonumber\\
	&=-\frac{1}{2}\Braket{P^{S}_{i_\text{out}}(\omega) P^C_{j _\text{out}}(\omega)+ P^C_{j_\text{out}} (\omega)  P^{S}_{i_\text{out}}( \omega) } \nonumber\\
	&= \frac{1}{4}\cbracket{- \Im{\sbracket{ B_{i_\text{out}}^- (\omega) B_{j_\text{out}}^- (\omega)^*+ A_{i_\text{out}}^- (\omega) A_{j_\text{out}}^- (\omega)^*}}- \rbracket{\omega \rightarrow -\omega}}\delta(0)
\end{align}

\begin{align}
	&\frac{1}{2}\Braket{Q^{C}_{i_\text{out}}(\omega) P^{C}_{j _\text{out}}(\omega)+ P^{C}_{j_\text{out}} (\omega)  Q^{C}_{i_\text{out}}( \omega) } \nonumber\\
	&=\frac{1}{2}\Braket{Q^{S}_{i_\text{out}}(\omega) P^{S}_{j _\text{out}}(\omega)+ P^{S}_{j_\text{out}} (\omega)  Q^{S}_{i_\text{out}}( \omega) } \nonumber\\
	&=\frac{1}{4}\cbracket{ -\Im{\sbracket{ B_{i_\text{out}}^+ (\omega) B_{j_\text{out}}^- (\omega)^*+ A_{i_\text{out}}^+ (\omega) A_{j_\text{out}}^- (\omega)^*}}+ \rbracket{\omega \rightarrow -\omega}}\delta(0)
\end{align}

\begin{align}
	&\frac{1}{2}\Braket{Q^{C}_{i_\text{out}}(\omega) P^{S}_{j _\text{out}}(\omega)+ P^{S}_{j_\text{out}} (\omega)  Q^{C}_{i_\text{out}}( \omega) } \nonumber\\
	&=-\frac{1}{2}\Braket{Q^{S}_{i_\text{out}}(\omega) P^{C}_{j _\text{out}}(\omega)+ P^{C}_{j_\text{out}} (\omega)  Q^{S}_{i_\text{out}}( \omega) } \nonumber\\
	&=\frac{1}{4}\cbracket{ -\Re{\sbracket{ B_{i_\text{out}}^+ (\omega) B_{j_\text{out}}^- (\omega)^*+ A_{i_\text{out}}^+ (\omega) A_{j_\text{out}}^- (\omega)^*}}- \rbracket{\omega \rightarrow -\omega}}\delta(0)
\end{align}

\begin{align}
	&\frac{1}{2}\Braket{P^{C}_{i_\text{out}}(\omega) Q^{C}_{j _\text{out}}(\omega)+ Q^{C}_{j_\text{out}} (\omega)  P^{C}_{i_\text{out}}( \omega) }\nonumber \\
	&=\frac{1}{2}\Braket{P^{S}_{i_\text{out}}(\omega) Q^{S}_{j _\text{out}}(\omega)+ Q^{S}_{j_\text{out}} (\omega)  P^{S}_{i_\text{out}}( \omega) }\nonumber \\
	&=\frac{1}{4}\cbracket{ \Im{\sbracket{ B_{i_\text{out}}^- (\omega) B_{j_\text{out}}^+ (\omega)^*+ A_{i_\text{out}}^- (\omega) A_{j_\text{out}}^+ (\omega)^*}}+ \rbracket{\omega \rightarrow -\omega}}\delta(0)
\end{align}

\begin{align}
	&\frac{1}{2}\Braket{P^{C}_{i_\text{out}}(\omega) Q^{S}_{j _\text{out}}(\omega)+ Q^{S}_{j_\text{out}} (\omega)  P^{C}_{i_\text{out}}( \omega) }\nonumber \\
	&=-\frac{1}{2}\Braket{P^{S}_{i_\text{out}}(\omega) Q^{C}_{j _\text{out}}(\omega)+ Q^{C}_{j_\text{out}} (\omega)  P^{S}_{i_\text{out}}( \omega) }\nonumber \\
	&=\frac{1}{4}\cbracket{ \Re{\sbracket{ B_{i_\text{out}}^- (\omega) B_{j_\text{out}}^+ (\omega)^*+ A_{i_\text{out}}^- (\omega) A_{j_\text{out}}^+ (\omega)^*}}- \rbracket{\omega \rightarrow -\omega}}\delta(0)
\end{align}

\section{Entropy and Quadrature Resonance Condition}

The resonance condition can be found be determining when the magnitude of the denominator of all elements of $T$ in \eqref{Tsol} 
are minimised. The matrix $T$ will have a prefactor of $1/\textrm{Det}\left(-i\omega I-A\right)$ and so the resonance condition will be determined by
\begin{align}
	\textrm{Det}\left(-i\omega I-A\right)=&\left(4\Delta^2+\left(\kappa-2 i\omega\right)^2\right)\left(\Gamma-2i\omega\right)^2-64 g_C^2\Delta\Omega_M+4\left(4\Delta^2+\left(\kappa-2i\omega\right)^2\right)\Omega_M^2 \nonumber\\
	=&\Bigg[4\gamma^2\Delta^2+\Gamma^2\kappa^2-4\Gamma^2\omega^2-16\Delta^2\omega^2 -16\Gamma\kappa\omega^2-4\kappa^2\omega^2+16\omega^4-64g_C^2\Delta\Omega_M\nonumber\\
	&+16\Delta^2\Omega_M^2+4\kappa^2\Omega_M^2-16\omega^2\Omega_M^2\Bigg] +4i\Bigg[4\Gamma\omega^3+4\kappa\omega^3-4\Gamma\kappa\omega^2-4\Gamma\Delta^2\omega\nonumber\\
	&-\Gamma^2\kappa\omega-\Gamma\kappa^2\omega-4\kappa\omega\Omega_M^2\Bigg]=0.
\end{align} 
By inspection, we find that the absolute value of the determinant is minimised at the resonance point, which corresponds to when the real part of the determinant is zero. Thus we only need to solve for
\begin{align}
	0=&\Bigg[4\Gamma^2\Delta^2+\Gamma^2\kappa^2-4\Gamma^2\omega^2-16\Delta^2\omega^2 -16\Gamma\kappa\omega^2-4\kappa^2\omega^2-64g_C^2\Delta\Omega_M\nonumber\\
	&+16\Delta^2\Omega_M^2+4\kappa^2\Omega_M^2-16\omega^2\Omega_M^2 +16\omega^4\Bigg]
\end{align}
This can be further simplified by grouping each term according to its functional dependence on $\omega$, numerically evaluating each prefactor and dropping all subdominant terms, leaving
\begin{equation}
	0=\alpha-\beta\omega^2+16\omega^4
\end{equation} 
with $\alpha=-64g_C^2\Delta\Omega_M$ and $\beta=16\Delta^2$. This admits the set of four solutions
\begin{equation}\label{4sol}
	\omega= \begin{cases} 
		\pm\frac{\sqrt{\beta+\sqrt{\beta^2-64\alpha}}}{4\sqrt{2}}  \\
		\pm\frac{\sqrt{\beta-\sqrt{\beta^2-64\alpha}}}{4\sqrt{2}}
	\end{cases}
\end{equation}

By numerically evaluating each solution, we determined that the relevant (real-valued) solutions are
the first two solutions in \eqref{4sol}, namely 
$\omega=\pm\frac{\sqrt{\beta-\sqrt{\beta^2-64\alpha}}}{4\sqrt{2}}\approx\pm\sqrt{\frac{\alpha}{\beta}}\approx\pm\sqrt{\frac{4\Omega_Mg_C^2}{-\Delta}}$. Note that the solutions are real as the detuning, $\Delta$, is negative. Also, at each step where an approximate appears only the leading order terms were kept. Due to the number of approximations used, the solution is only accurate to the first order and does not match the scaling seen at very small $\tilde{P}$.
\end{widetext}


\begin{thebibliography}{25}%
	\makeatletter
	\providecommand \@ifxundefined [1]{%
		\@ifx{#1\undefined}
	}%
	\providecommand \@ifnum [1]{%
		\ifnum #1\expandafter \@firstoftwo
		\else \expandafter \@secondoftwo
		\fi
	}%
	\providecommand \@ifx [1]{%
		\ifx #1\expandafter \@firstoftwo
		\else \expandafter \@secondoftwo
		\fi
	}%
	\providecommand \natexlab [1]{#1}%
	\providecommand \enquote  [1]{``#1''}%
	\providecommand \bibnamefont  [1]{#1}%
	\providecommand \bibfnamefont [1]{#1}%
	\providecommand \citenamefont [1]{#1}%
	\providecommand \href@noop [0]{\@secondoftwo}%
	\providecommand \href [0]{\begingroup \@sanitize@url \@href}%
	\providecommand \@href[1]{\@@startlink{#1}\@@href}%
	\providecommand \@@href[1]{\endgroup#1\@@endlink}%
	\providecommand \@sanitize@url [0]{\catcode `\\12\catcode `\$12\catcode
		`\&12\catcode `\#12\catcode `\^12\catcode `\_12\catcode `\%12\relax}%
	\providecommand \@@startlink[1]{}%
	\providecommand \@@endlink[0]{}%
	\providecommand \url  [0]{\begingroup\@sanitize@url \@url }%
	\providecommand \@url [1]{\endgroup\@href {#1}{\urlprefix }}%
	\providecommand \urlprefix  [0]{URL }%
	\providecommand \Eprint [0]{\href }%
	\providecommand \doibase [0]{http://dx.doi.org/}%
	\providecommand \selectlanguage [0]{\@gobble}%
	\providecommand \bibinfo  [0]{\@secondoftwo}%
	\providecommand \bibfield  [0]{\@secondoftwo}%
	\providecommand \translation [1]{[#1]}%
	\providecommand \BibitemOpen [0]{}%
	\providecommand \bibitemStop [0]{}%
	\providecommand \bibitemNoStop [0]{.\EOS\space}%
	\providecommand \EOS [0]{\spacefactor3000\relax}%
	\providecommand \BibitemShut  [1]{\csname bibitem#1\endcsname}%
	\let\auto@bib@innerbib\@empty
	
	\bibitem [{\citenamefont {Marino}\ \emph {et~al.}(2010)\citenamefont {Marino},
		\citenamefont {Cataliotti}, \citenamefont {Farsi}, \citenamefont {de~Cumis},\
		and\ \citenamefont {Marin}}]{marino_classical_2010}%
	\BibitemOpen
	\bibfield  {author} {\bibinfo {author} {\bibfnamefont {F.}~\bibnamefont
			{Marino}}, \bibinfo {author} {\bibfnamefont {F.~S.}\ \bibnamefont
			{Cataliotti}}, \bibinfo {author} {\bibfnamefont {A.}~\bibnamefont {Farsi}},
		\bibinfo {author} {\bibfnamefont {M.~S.}\ \bibnamefont {de~Cumis}}, \ and\
		\bibinfo {author} {\bibfnamefont {F.}~\bibnamefont {Marin}},\ }\href
	{\doibase 10.1103/PhysRevLett.104.073601} {\bibfield  {journal} {\bibinfo
			{journal} {Physical Review Letters}\ }\textbf {\bibinfo {volume} {104}},\
		\bibinfo {pages} {073601} (\bibinfo {year} {2010})}\BibitemShut {NoStop}%
	\bibitem [{\citenamefont {Safavi-Naeini}\ \emph {et~al.}(2013)\citenamefont
		{Safavi-Naeini}, \citenamefont {Gr{\''o}blacher}, \citenamefont {Hill},
		\citenamefont {Chan}, \citenamefont {Aspelmeyer},\ and\ \citenamefont
		{Painter}}]{safavi-naeini_squeezed_2013}%
	\BibitemOpen
	\bibfield  {author} {\bibinfo {author} {\bibfnamefont {A.~H.}\ \bibnamefont
			{Safavi-Naeini}}, \bibinfo {author} {\bibfnamefont {S.}~\bibnamefont
			{Gr{\''o}blacher}}, \bibinfo {author} {\bibfnamefont {J.~T.}\ \bibnamefont
			{Hill}}, \bibinfo {author} {\bibfnamefont {J.}~\bibnamefont {Chan}}, \bibinfo
		{author} {\bibfnamefont {M.}~\bibnamefont {Aspelmeyer}}, \ and\ \bibinfo
		{author} {\bibfnamefont {O.}~\bibnamefont {Painter}},\ }\href {\doibase
		10.1038/nature12307} {\bibfield  {journal} {\bibinfo  {journal} {Nature}\
		}\textbf {\bibinfo {volume} {500}},\ \bibinfo {pages} {185} (\bibinfo {year}
		{2013})}\BibitemShut {NoStop}%
	\bibitem [{\citenamefont {Gr\"{o}blacher}\ \emph {et~al.}(2009)\citenamefont
		{Gr\"{o}blacher}, \citenamefont {Hertzberg}, \citenamefont {Vanner},
		\citenamefont {Cole}, \citenamefont {Gigan}, \citenamefont {Schwab},\ and\
		\citenamefont {Aspelmeyer}}]{groblacher_demonstration_2009}%
	\BibitemOpen
	\bibfield  {author} {\bibinfo {author} {\bibfnamefont {S.}~\bibnamefont
			{Gr\"{o}blacher}}, \bibinfo {author} {\bibfnamefont {J.~B.}\ \bibnamefont
			{Hertzberg}}, \bibinfo {author} {\bibfnamefont {M.~R.}\ \bibnamefont
			{Vanner}}, \bibinfo {author} {\bibfnamefont {G.~D.}\ \bibnamefont {Cole}},
		\bibinfo {author} {\bibfnamefont {S.}~\bibnamefont {Gigan}}, \bibinfo
		{author} {\bibfnamefont {K.~C.}\ \bibnamefont {Schwab}}, \ and\ \bibinfo
		{author} {\bibfnamefont {M.}~\bibnamefont {Aspelmeyer}},\ }\href {\doibase
		10.1038/nphys1301} {\bibfield  {journal} {\bibinfo  {journal} {Nature
				Physics}\ }\textbf {\bibinfo {volume} {5}},\ \bibinfo {pages} {485} (\bibinfo
		{year} {2009})}\BibitemShut {NoStop}%
	\bibitem [{\citenamefont {Tsang}(2013)}]{tsang_quantum_2013}%
	\BibitemOpen
	\bibfield  {author} {\bibinfo {author} {\bibfnamefont {M.}~\bibnamefont
			{Tsang}},\ }\href {\doibase 10.1088/1367-2630/15/7/073005} {\bibfield
		{journal} {\bibinfo  {journal} {New Journal of Physics}\ }\textbf {\bibinfo
			{volume} {15}},\ \bibinfo {pages} {073005} (\bibinfo {year}
		{2013})}\BibitemShut {NoStop}%
	\bibitem [{\citenamefont {Carmon}\ \emph {et~al.}(2005)\citenamefont {Carmon},
		\citenamefont {Rokhsari}, \citenamefont {Yang}, \citenamefont {Kippenberg},\
		and\ \citenamefont {Vahala}}]{carmon_temporal_2005}%
	\BibitemOpen
	\bibfield  {author} {\bibinfo {author} {\bibfnamefont {T.}~\bibnamefont
			{Carmon}}, \bibinfo {author} {\bibfnamefont {H.}~\bibnamefont {Rokhsari}},
		\bibinfo {author} {\bibfnamefont {L.}~\bibnamefont {Yang}}, \bibinfo {author}
		{\bibfnamefont {T.~J.}\ \bibnamefont {Kippenberg}}, \ and\ \bibinfo {author}
		{\bibfnamefont {K.~J.}\ \bibnamefont {Vahala}},\ }\href {\doibase
		10.1103/PhysRevLett.94.223902} {\bibfield  {journal} {\bibinfo  {journal}
			{Physical Review Letters}\ }\textbf {\bibinfo {volume} {94}},\ \bibinfo
		{pages} {223902} (\bibinfo {year} {2005})}\BibitemShut {NoStop}%
	\bibitem [{\citenamefont {Sun}\ \emph {et~al.}(2012)\citenamefont {Sun},
		\citenamefont {Zheng}, \citenamefont {Poot}, \citenamefont {Wong},\ and\
		\citenamefont {Tang}}]{sun_femtogram_2012}%
	\BibitemOpen
	\bibfield  {author} {\bibinfo {author} {\bibfnamefont {X.}~\bibnamefont
			{Sun}}, \bibinfo {author} {\bibfnamefont {J.}~\bibnamefont {Zheng}}, \bibinfo
		{author} {\bibfnamefont {M.}~\bibnamefont {Poot}}, \bibinfo {author}
		{\bibfnamefont {C.~W.}\ \bibnamefont {Wong}}, \ and\ \bibinfo {author}
		{\bibfnamefont {H.~X.}\ \bibnamefont {Tang}},\ }\href {\doibase
		10.1021/nl300142t} {\bibfield  {journal} {\bibinfo  {journal} {Nano Letters}\
		}\textbf {\bibinfo {volume} {12}},\ \bibinfo {pages} {2299} (\bibinfo {year}
		{2012})}\BibitemShut {NoStop}%
	\bibitem [{\citenamefont {Corbitt}\ \emph {et~al.}(2006)\citenamefont
		{Corbitt}, \citenamefont {Ottaway}, \citenamefont {Innerhofer}, \citenamefont
		{Pelc},\ and\ \citenamefont {Mavalvala}}]{corbitt_measurement_2006}%
	\BibitemOpen
	\bibfield  {author} {\bibinfo {author} {\bibfnamefont {T.}~\bibnamefont
			{Corbitt}}, \bibinfo {author} {\bibfnamefont {D.}~\bibnamefont {Ottaway}},
		\bibinfo {author} {\bibfnamefont {E.}~\bibnamefont {Innerhofer}}, \bibinfo
		{author} {\bibfnamefont {J.}~\bibnamefont {Pelc}}, \ and\ \bibinfo {author}
		{\bibfnamefont {N.}~\bibnamefont {Mavalvala}},\ }\href {\doibase
		10.1103/PhysRevA.74.021802} {\bibfield  {journal} {\bibinfo  {journal}
			{Physical Review A}\ }\textbf {\bibinfo {volume} {74}},\ \bibinfo {pages}
		{021802} (\bibinfo {year} {2006})}\BibitemShut {NoStop}%
	\bibitem [{\citenamefont {Corbitt}\ and\ \citenamefont
		{Mavalvala}()}]{corbitt_review:_2004}%
	\BibitemOpen
	\bibfield  {author} {\bibinfo {author} {\bibfnamefont {T.}~\bibnamefont
			{Corbitt}}\ and\ \bibinfo {author} {\bibfnamefont {N.}~\bibnamefont
			{Mavalvala}},\ }\href {\doibase 10.1088/1464-4266/6/8/008} {\ \textbf
		{\bibinfo {volume} {6}},\ \bibinfo {pages} {S675}}\BibitemShut {NoStop}%
	\bibitem [{\citenamefont {Gro{\ss}ardt}\ \emph {et~al.}(2016)\citenamefont
		{Gro{\ss}ardt}, \citenamefont {Bateman}, \citenamefont {Ulbricht},\ and\
		\citenamefont {Bassi}}]{grosardt_optomechanical_2016}%
	\BibitemOpen
	\bibfield  {author} {\bibinfo {author} {\bibfnamefont {A.}~\bibnamefont
			{Gro{\ss}ardt}}, \bibinfo {author} {\bibfnamefont {J.}~\bibnamefont
			{Bateman}}, \bibinfo {author} {\bibfnamefont {H.}~\bibnamefont {Ulbricht}}, \
		and\ \bibinfo {author} {\bibfnamefont {A.}~\bibnamefont {Bassi}},\ }\href
	{\doibase 10.1103/PhysRevD.93.096003} {\bibfield  {journal} {\bibinfo
			{journal} {Physical Review D}\ }\textbf {\bibinfo {volume} {93}},\ \bibinfo
		{pages} {096003} (\bibinfo {year} {2016})}\BibitemShut {NoStop}%
	\bibitem [{\citenamefont {Gan}\ \emph {et~al.}(2016)\citenamefont {Gan},
		\citenamefont {Savage},\ and\ \citenamefont
		{Scully}}]{gan_optomechanical_2016}%
	\BibitemOpen
	\bibfield  {author} {\bibinfo {author} {\bibfnamefont {C.}~\bibnamefont
			{Gan}}, \bibinfo {author} {\bibfnamefont {C.}~\bibnamefont {Savage}}, \ and\
		\bibinfo {author} {\bibfnamefont {S.}~\bibnamefont {Scully}},\ }\href
	{\doibase 10.1103/PhysRevD.93.124049} {\bibfield  {journal} {\bibinfo
			{journal} {Physical Review D}\ }\textbf {\bibinfo {volume} {93}},\ \bibinfo
		{pages} {124049} (\bibinfo {year} {2016})}\BibitemShut {NoStop}%
	\bibitem [{\citenamefont {Arvanitaki}\ and\ \citenamefont
		{Geraci}(2013)}]{arvanitaki_detecting_2013}%
	\BibitemOpen
	\bibfield  {author} {\bibinfo {author} {\bibfnamefont {A.}~\bibnamefont
			{Arvanitaki}}\ and\ \bibinfo {author} {\bibfnamefont {A.~A.}\ \bibnamefont
			{Geraci}},\ }\href {\doibase 10.1103/PhysRevLett.110.071105} {\bibfield
		{journal} {\bibinfo  {journal} {Physical Review Letters}\ }\textbf {\bibinfo
			{volume} {110}},\ \bibinfo {pages} {071105} (\bibinfo {year}
		{2013})}\BibitemShut {NoStop}%
	\bibitem [{\citenamefont {Libbrecht}\ and\ \citenamefont
		{Black}(2004)}]{libbrecht_toward_2004}%
	\BibitemOpen
	\bibfield  {author} {\bibinfo {author} {\bibfnamefont {K.~G.}\ \bibnamefont
			{Libbrecht}}\ and\ \bibinfo {author} {\bibfnamefont {E.~D.}\ \bibnamefont
			{Black}},\ }\href {\doibase 10.1016/j.physleta.2003.12.022} {\bibfield
		{journal} {\bibinfo  {journal} {Physics Letters A}\ }\textbf {\bibinfo
			{volume} {321}},\ \bibinfo {pages} {99} (\bibinfo {year} {2004})}\BibitemShut
	{NoStop}%
	\bibitem [{\citenamefont {Guccione}\ \emph {et~al.}(2013)\citenamefont
		{Guccione}, \citenamefont {Hosseini}, \citenamefont {Adlong}, \citenamefont
		{Johnsson}, \citenamefont {Hope}, \citenamefont {Buchler},\ and\
		\citenamefont {Lam}}]{guccione_scattering-free_2013}%
	\BibitemOpen
	\bibfield  {author} {\bibinfo {author} {\bibfnamefont {G.}~\bibnamefont
			{Guccione}}, \bibinfo {author} {\bibfnamefont {M.}~\bibnamefont {Hosseini}},
		\bibinfo {author} {\bibfnamefont {S.}~\bibnamefont {Adlong}}, \bibinfo
		{author} {\bibfnamefont {M.~T.}\ \bibnamefont {Johnsson}}, \bibinfo {author}
		{\bibfnamefont {J.}~\bibnamefont {Hope}}, \bibinfo {author} {\bibfnamefont
			{B.~C.}\ \bibnamefont {Buchler}}, \ and\ \bibinfo {author} {\bibfnamefont
			{P.~K.}\ \bibnamefont {Lam}},\ }\href {\doibase
		10.1103/PhysRevLett.111.183001} {\bibfield  {journal} {\bibinfo  {journal}
			{Physical Review Letters}\ }\textbf {\bibinfo {volume} {111}},\ \bibinfo
		{pages} {183001} (\bibinfo {year} {2013})}\BibitemShut {NoStop}%
	\bibitem [{\citenamefont {Michimura}\ \emph {et~al.}(2017)\citenamefont
		{Michimura}, \citenamefont {Kuwahara}, \citenamefont {Ushiba}, \citenamefont
		{Matsumoto},\ and\ \citenamefont {Ando}}]{michimura_optical_2017}%
	\BibitemOpen
	\bibfield  {author} {\bibinfo {author} {\bibfnamefont {Y.}~\bibnamefont
			{Michimura}}, \bibinfo {author} {\bibfnamefont {Y.}~\bibnamefont {Kuwahara}},
		\bibinfo {author} {\bibfnamefont {T.}~\bibnamefont {Ushiba}}, \bibinfo
		{author} {\bibfnamefont {N.}~\bibnamefont {Matsumoto}}, \ and\ \bibinfo
		{author} {\bibfnamefont {M.}~\bibnamefont {Ando}},\ }\href {\doibase
		10.1364/OE.25.013799} {\bibfield  {journal} {\bibinfo  {journal} {Optics
				Express}\ }\textbf {\bibinfo {volume} {25}},\ \bibinfo {pages} {13799}
		(\bibinfo {year} {2017})}\BibitemShut {NoStop}%
\bibitem {SIN10} S. Singh, G. A. Phelps, D. S. Goldbaum, E. M. Wright, and P. Meystre, Phys.Rev.Lett. {\bf 105}, 213602 (2010).
		\bibitem {ISA10} Oriol Romero-Isart et al., New J. Phys. {\bf 12}, 033015 (2010).
		\bibitem {CHA10} D. E. Chang
et al., Proc. Natl. Acad. Sci. U.S.A. {\bf 107}, 1005 (2010).	

	\bibitem [{\citenamefont {Bowen}\ and\ \citenamefont
		{Milburn}(2015)}]{bowen_quantum_2015}%
	\BibitemOpen
	\bibfield  {author} {\bibinfo {author} {\bibfnamefont {W.~P.}\ \bibnamefont
			{Bowen}}\ and\ \bibinfo {author} {\bibfnamefont {G.~J.}\ \bibnamefont
			{Milburn}},\ }\href@noop {} {\emph {\bibinfo {title} {Quantum
				Optomechanics}}}\ (\bibinfo  {publisher} {{CRC} Press},\ \bibinfo {address}
	{Boca Raton. {FL}},\ \bibinfo {year} {2015})\BibitemShut {NoStop}%
	\bibitem [{\citenamefont {Gardiner}\ and\ \citenamefont
		{Collett}(1985)}]{gardiner_input_1985}%
	\BibitemOpen
	\bibfield  {author} {\bibinfo {author} {\bibfnamefont {C.~W.}\ \bibnamefont
			{Gardiner}}\ and\ \bibinfo {author} {\bibfnamefont {M.~J.}\ \bibnamefont
			{Collett}},\ }\href {\doibase 10.1103/PhysRevA.31.3761} {\bibfield  {journal}
		{\bibinfo  {journal} {Physical Review A}\ }\textbf {\bibinfo {volume} {31}},\
		\bibinfo {pages} {3761} (\bibinfo {year} {1985})}\BibitemShut {NoStop}%
	\bibitem [{Note1()}]{Note1}%
	\BibitemOpen
	\bibinfo {note} {Due to the requirement that the linearised part is small,
		$\protect \Braket {\protect \mathaccentV {hat}05E{a}^\dagger \protect
			\mathaccentV {hat}05E{a}} = \protect \Braket {\protect \mathaccentV
			{hat}05E{a}^\dagger }\protect \Braket {\protect \mathaccentV {hat}05E{a}} +
		\protect \Braket {\delta \protect \mathaccentV {hat}05E{a}^\dagger \delta
			\protect \mathaccentV {hat}05E{a}} \approx \protect \Braket {\protect
			\mathaccentV {hat}05E{a}^\dagger }\protect \Braket {\protect \mathaccentV
			{hat}05E{a}}$.}\BibitemShut {Stop}%
	\bibitem [{Note2()}]{Note2}%
	\BibitemOpen
	\bibinfo {note} {$N_\protect \text {in}$ has dimensions of photons per
		second.}\BibitemShut {Stop}%
	\bibitem [{Note3()}]{Note3}%
	\BibitemOpen
	\bibinfo {note} {See Supplemental Material at [URL will be inserted by
		publisher] for more details.}\BibitemShut {Stop}%
	\bibitem [{\citenamefont {Schumaker}(1986)}]{schumaker_quantum_1986}%
	\BibitemOpen
	\bibfield  {author} {\bibinfo {author} {\bibfnamefont {B.~L.}\ \bibnamefont
			{Schumaker}},\ }\href {\doibase 10.1016/0370-1573(86)90179-1} {\bibfield
		{journal} {\bibinfo  {journal} {Physics Reports}\ }\textbf {\bibinfo {volume}
			{135}},\ \bibinfo {pages} {317} (\bibinfo {year} {1986})}\BibitemShut
	{NoStop}%
	\bibitem [{\citenamefont {Adesso}\ \emph {et~al.}(2014)\citenamefont {Adesso},
		\citenamefont {Ragy},\ and\ \citenamefont {Lee}}]{adesso_continuous_2014}%
	\BibitemOpen
	\bibfield  {author} {\bibinfo {author} {\bibfnamefont {G.}~\bibnamefont
			{Adesso}}, \bibinfo {author} {\bibfnamefont {S.}~\bibnamefont {Ragy}}, \ and\
		\bibinfo {author} {\bibfnamefont {A.~R.}\ \bibnamefont {Lee}},\ }\href
	{\doibase 10.1142/S1230161214400010} {\bibfield  {journal} {\bibinfo
			{journal} {Open Systems \& Information Dynamics}\ }\textbf {\bibinfo {volume}
			{21}},\ \bibinfo {pages} {1440001} (\bibinfo {year} {2014})}\BibitemShut
	{NoStop}%
	\bibitem [{\citenamefont {Ralph}\ \emph {et~al.}(2008)\citenamefont {Ralph},
		\citenamefont {Huntington},\ and\ \citenamefont
		{Symul}}]{ralph_single-photon_2008}%
	\BibitemOpen
	\bibfield  {author} {\bibinfo {author} {\bibfnamefont {T.~C.}\ \bibnamefont
			{Ralph}}, \bibinfo {author} {\bibfnamefont {E.~H.}\ \bibnamefont
			{Huntington}}, \ and\ \bibinfo {author} {\bibfnamefont {T.}~\bibnamefont
			{Symul}},\ }\href {\doibase 10.1103/PhysRevA.77.063817} {\bibfield  {journal}
		{\bibinfo  {journal} {Physical Review A}\ }\textbf {\bibinfo {volume} {77}},\
		\bibinfo {pages} {063817} (\bibinfo {year} {2008})}\BibitemShut {NoStop}%
	\bibitem [{\citenamefont {Huan}\ \emph {et~al.}(2015)\citenamefont {Huan},
		\citenamefont {Zhou},\ and\ \citenamefont {Ian}}]{PhysRevA.92.022301}%
	\BibitemOpen
	\bibfield  {author} {\bibinfo {author} {\bibfnamefont {T.}~\bibnamefont
			{Huan}}, \bibinfo {author} {\bibfnamefont {R.}~\bibnamefont {Zhou}}, \ and\
		\bibinfo {author} {\bibfnamefont {H.}~\bibnamefont {Ian}},\ }\href {\doibase
		10.1103/PhysRevA.92.022301} {\bibfield  {journal} {\bibinfo  {journal} {Phys.
				Rev. A}\ }\textbf {\bibinfo {volume} {92}},\ \bibinfo {pages} {022301}
		(\bibinfo {year} {2015})}\BibitemShut {NoStop}%
	\bibitem [{\citenamefont {Vanner}\ \emph {et~al.}(2013)\citenamefont {Vanner},
		\citenamefont {Hofer}, \citenamefont {Cole},\ and\ \citenamefont
		{Aspelmeyer}}]{vanner_cooling-by-measurement_2013}%
	\BibitemOpen
	\bibfield  {author} {\bibinfo {author} {\bibfnamefont {M.~R.}\ \bibnamefont
			{Vanner}}, \bibinfo {author} {\bibfnamefont {J.}~\bibnamefont {Hofer}},
		\bibinfo {author} {\bibfnamefont {G.~D.}\ \bibnamefont {Cole}}, \ and\
		\bibinfo {author} {\bibfnamefont {M.}~\bibnamefont {Aspelmeyer}},\ }\href
	{https://doi.org/10.1038/ncomms3295} {\bibfield  {journal} {\bibinfo
			{journal} {Nature Communications}\ }\textbf {\bibinfo {volume} {4}},\
		\bibinfo {pages} {2295} (\bibinfo {year} {2013})}\BibitemShut {NoStop}%
\end{thebibliography}
%

\end{document}